\RequirePackage{fix-cm}
\documentclass[twocolumn,epjc3]{svjour3}  

\smartqed 
\RequirePackage{graphicx}
\usepackage[usenames,dvipsnames]{xcolor}
\usepackage{graphicx}
\usepackage{dcolumn}
\usepackage{bm}
\usepackage{hyperref}
\usepackage{lipsum}
\usepackage[normalem]{ulem}
\usepackage{url}

\usepackage{siunitx}

\renewcommand{\Re}{\text{Re}}

\usepackage{amsmath,eqnarray}
\DeclareMathOperator{\Tr}{Tr}
\DeclareMathOperator{\tr}{tr}

\usepackage{amssymb}

\usepackage{dsfont}

\usepackage[ruled,vlined]{algorithm2e}

\journalname{Eur. Phys. J. C}

\begin{document}

\allowdisplaybreaks

\title{Finite-temperature mean-field approximations for shell model 
       Hamiltonians: the code HF-SHELL}
\author{W. Ryssens \thanksref{addr1,e1}
        \and
        Y. Alhassid \thanksref{addr1,e2} 
}
\thankstext{e1}{e-mail: wouter.ryssens@yale.edu, Current address:
               Institut d'Astronomie et d'Astrophysique, 
               Universit\'e Libre de Bruxelles, CP 226, Brussels 1050, Belgium  }
\thankstext{e2}{e-mail: yoram.alhassid@yale.edu (corresponding author)}
\institute{Center for Theoretical Physics, 
             Sloane Physics Laboratory,
             Yale University, New Haven, Connecticut 06520, USA \label{addr1}}
\date{Received: date / Accepted: date}

\maketitle

\begin{abstract}
We present the code HF-SHELL for solving the self-consistent mean-field equations
for configuration-interaction shell model Hamiltonians in the proton-neutron 
formalism. The code can calculate both zero- and finite-temperature 
properties in the Hartree-Fock (HF), HF + Bardeen-Cooper-Schrieffer (HF+BCS) and the
Hartree-Fock-Bogoliubov (HFB) mean-field approximations. \\Particle-number 
projection after variation is incorporated to reduce the grand-canonical 
ensemble to the canonical ensemble, making the code particularly suitable 
for the calculation of nuclear state densities. The code does not impose axial 
symmetry and allows for triaxial quadrupole deformations. The 
self-consistency cycle is particularly robust through the use of the heavy-ball 
optimization technique and the implementation of different options to constrain 
the quadrupole degrees of freedom.

\end{abstract}

\section*{PROGRAM SUMMARY}
\begin{description}[font=\normalfont\itshape]
  \item[\em Program title:] HF-SHELL
  \item[\em Licensing provisions:] GNU General Public License Version 3 or later
  \item[\em Programming language:] Fortran
  \item[\em Repository and DOI:]  
    \href{https://github.com/wryssens/hf-shell}{\nolinkurl{github.com/wryssens/hf-shell}} \\
       { DOI: \href{https://doi.org/10.5281/zenodo.4008440}{\nolinkurl{10.5281/zenodo.4008440}}}
  \item[\em Description of problem:] 
    HF-SHELL solves the self-consistent mean-field problem {at finite temperature} for nuclear 
    shell-model Hamiltonians on the level of the Hartree-Fock (HF), the 
    HF-Bogoliubov (HFB) or the HF+Bardeen-Cooper-Schrieffer (HF+BCS) 
    approximations. {The code also includes the option for
    zero-temperature calculations.}
    The particle-number projection after variation method of Ref.~\cite{Fanto2017} 
    is implemented to reduce the grand-canonical ensemble of mean-field theory 
    to the canonical ensemble. The code can be used to calculate nuclear 
    state densities and to generate free energy surfaces at finite temperature
    as a function of the nuclear shape. 
  \item[\em Solution method:]
    The HF basis is iterated through the heavy-ball algorithm of Ref.~\cite{Ryssens2019}, 
    and is supplemented by the two-basis method~\cite{Gall1994} in the case of the HFB approximation. 
  \item[\em Additional comments:] 
      { The code is fast:
        for the examples provided, a single calculation on a laptop takes from 
        much less than a second for $^{24}$Mg up to a few seconds for $^{162}$Dy.
        A complete calculation of the state 
        density of $^{162}$Dy as a function of the excitation energy (involving  
        roughly 500 individual mean-field calculations) takes about ten minutes. 
        The code uses only a small amount of memory requiring less than  300 MB 
        of memory in all the examples shown.}   
        Several {symmetry} restrictions are imposed on the allowed nuclear 
        configuration: time reversal, z-signature and reflection symmetries. 
        In addition, the matrix elements of the relevant mean-field transformation 
        (for HF, HFB or HF+BCS) are assumed to be real. 
        The model space and effective interaction should be specified in the 
        proton-neutron formalism, in which protons and neutrons can occupy 
        different valence {shells}.
\end{description}

\section{Introduction}

{Self-consistent} mean-field approximations are an important tool in nuclear
 physics~\cite{Bender03}. Due to their computational simplicity, they can be 
applied across the nuclear chart, but they often miss important many-body 
correlations. A partial solution to this deficiency is to allow for spontaneous 
symmetry breaking.  Symmetry breaking lowers the energy of the mean-field 
solution, but results in the loss of good quantum numbers, making a direct 
comparison between theory and experimental results difficult. On the other hand,
spontaneous breaking of symmetries offers an intuitive picture of various 
aspects of nuclear structure. For example, intrinsic quadrupole deformation 
arises naturally in a mean-field formalism that breaks rotational symmetry, and 
has been a cornerstone in our understanding of heavy nuclei. Because of their 
computational simplicity and the physical interpretation of symmetry breaking, 
mean-field methods remain widely used. 

{More advanced many-body methods are generally limited by their high 
computational cost and are usually limited to relatively small model spaces. 
In such model spaces where advanced many-body methods are feasible, mean-field 
calculations are usually not competitive in terms 
of approximating the solution to the many-body problem.  Yet mean-field configurations are of interest 
in that they provide a benchmark and allow for 
physically valuable insight into more complex many-body calculations. 
Their computational simplicity offers an advantage even in small model spaces when carrying out  a sequence of many-body calculations become 
prohibitively costly.
Mean-field configurations are also useful as the starting point of numerous many-body methods, such as 
the  random-phase approximation,  symmetry restoration
methods~\cite{RingSchuck}, and the Monte Carlo Shell Model method~\cite{Otsuka2001}.}

The configuration-interaction (CI) shell model is {one such advanced} 
many-body method. {It provides an exact solution} for a given effective 
Hamiltonian in a valence model space. Direct diagonalization methods can be 
applied only in relatively small model spaces, while the shell model Monte Carlo
 (SMMC) method enables calculations in model spaces that are many 
orders of magnitude larger~\cite{Alhassid2017}. {SMMC calculations are 
computationally intensive and have not been applied to large
regions across the nuclear chart.} Mean-field calculations within the same CI 
shell-model space {are much faster and} provide insight into the interpretation 
of CI shell-model results; see, e.g., 
Refs.~\cite{Gao2015,Alhassid2016,Bally2019}. 

Aside from describing ground-state properties, mean-field methods have also 
been applied to describe nuclear properties at finite excitation energy, especially
in the context of compound-nucleus reactions. Since the compound nucleus 
equilibrates on a time scale that is short compared to its decay, a commonly 
used method is a finite-temperature mean-field theory, see, e.g., 
Ref.~\cite{Schunck2015} for {an application to} induced fission.  

One of the most important statistical nuclear properties is the nuclear level 
density (NLD), {which is a necessary input to the statistical theory
of nuclear reactions.}
{NLD models are often based on mean-field approximations, 
either directly using the mean-field partition function as in 
Ref.~\cite{Martin2003}, or by providing a single-particle spectrum for 
combinatorial models as in Ref.~\cite{Hilaire2012}. These models are
often global, especially when based on energy density functionals, but they
have to be augmented by phenomenological modeling of collective effects, 
such as the so-called ``rotational correction"~\cite{Bjornholm73}. }
{ The SMMC method on the other hand, provides exact NLDs 
(up to statistical errors) that already include the effects of  
collectivity. Its requires, however, suitable effective interactions in different mass regions.}
{ Mean-field densities count only the intrinsic states, and the ratio of the 
SMMC density to the mean-field density provides a microscopic estimate of the 
NLD enhancement due to collective states built on top of intrinsic 
states~\cite{Alhassid2016}. The understanding of this collective enhancement 
can lead to an improved treatment of collective corrections in global mean-field
based NLD models.
}

Many zero-temperature mean-field codes have been developed.  Most publicly 
available codes are designed {for} use with energy density functionals; see, 
e.g., Refs.~\cite{Maruhn2014,Niksic2014,Ryssens2015a,Perez2017,Schunck2017} and 
many others. {A zero-temperature code that performs variation after 
particle-number projection and is suitable for CI shell model Hamiltonians, was 
recently introduced in Ref.~\cite{Bally2020}.}
Published codes suitable for other types of effective nuclear 
interactions also exist but are less common; see, e.g., 
Refs.~\cite{Dobrowolski2016,Garcia1999}. However, little is available for 
finite-temperature mean-field approaches. To our knowledge HFBTHO~\cite{Stoitsov2013} and
HFODD~\cite{Schunck2012} are the only published energy density functional codes 
that allow for finite-temperature calculations. The code HFGRAD~\cite{Bertsch2016} 
is the only published code for use with CI shell-model Hamiltonians, and it 
includes Hartree-Fock (HF) mean-field routines for both zero- and 
finite-temperature properties. The finite-temperature mean-field code developed 
previously and used in Refs.~\cite{Alhassid2016,Ozen2013} is unpublished but 
offers comparable functionality and also includes the Hartree-Fock-Bogoliubov 
(HFB) mean-field approximation.

To address this lack of published finite-temperature mean-field codes for shell 
model Hamiltonians, we present here the code HF-SHELL, 
which solves the self-consistent mean-field equations for CI shell-model 
Hamiltonians both at zero and at finite temperature. {Its required computational
resources are relatively low, with runtimes on the order of seconds and 
memory usage of a few hundred MB}. As described below, {HF-SHELL} 
generalizes substantially the codes of Refs.~\cite{Alhassid2016} and \cite{Bertsch2016},
enabling the treatment of pairing correlations, triaxial deformations and the 
use of quadrupole constraints at finite temperature. The code is particularly 
suited for the calculation of nuclear level densities as it includes the 
particle-number projection techniques developed in Ref.~\cite{Fanto2017}. 
Numerically, the code offers a robust self-consistency cycle through the use of 
the heavy-ball method of Ref.~\cite{Ryssens2019}, which is similar in spirit to 
the gradient method of Refs.~\cite{Robledo2011,Bertsch2016}.

This paper is organized as follows:  in Sec.~\ref{sec:mean-field} we summarize 
the mean-field equations at finite temperature for the HF,  the HF plus
Bardeen-Cooper-Schrieffer (BCS) and the HFB approximations. 
In Sec.~\ref{sec:levelden} we describe the calculation of nuclear state
densities by using the ensemble reduction techniques of Ref.~\cite{Fanto2017}.
 In Sec.~\ref{sec:observables} we discuss several 
other observables that are calculated by the code, and in 
Sec.~\ref{sec:numerical} we describe the numerical techniques used to iterate and 
find the self-consistent solution. {In Sec.~\ref{sec:examples} we present 
several examples that demonstrate results obtained by the code, and in 
Sec.~\ref{sec:usage} we explain how to use the code, and describe in detail its 
input and output. We conclude in Sec.~\ref{sec:conclusion}.}

\section{Finite-temperature mean-field approximations}
\label{sec:mean-field}

The zero-temperature mean-field equations are  
well documented in the literature and textbooks~\cite{RingSchuck}.  However, 
their extension to finite temperature is not as well discussed in the literature
and we will briefly review key aspects of the formalism, following the original 
presentation in Ref.~\cite{Goodman1981}. For more details, we 
refer the reader to Refs.~\cite{BlaizotRipka} and \cite{Schunck2019}. 

While the zero-temperature formalism is based on pure states, the 
finite-temperature mean-field approximation is formulated in terms of a 
statistical mixture or density matrix $\hat D$. In the following we assume that 
$\hat D$ is normalized, i.e., $\Tr \hat D =1$.  
{The thermal expectation values of an observable $\hat{O}$ can be calculated from
\begin{align}
\langle \hat{O} \rangle_T = \Tr \left( \hat{D} \hat{O} \right) \,,
\end{align}
where the trace is taken over Fock space.}
At a given temperature $T$ and 
chemical potentials $\mu_p$ and $\mu_n$ (for protons and neutrons, respectively), we minimize 
the grand-canonical potential $\Omega$
\begin{equation}
\Omega(T,\mu_p,\mu_n) = E - T S - \mu_p N_p - \mu_n N_n \;,
\end{equation}
where the energy $E$, entropy $S$ and average particle numbers $N_q$ 
($q=p,n$) are given by
\begin{subequations}
\begin{align}
E   =&  \Tr \left( \hat{D} \hat{H}     \right) \, , \\
S   =& -\Tr \left( \hat{D} \ln \hat{D} \right) \, , \\
N_q =&  \Tr \left( \hat{D} \hat{N}_q   \right) \;,
\label{eq:E-S-N} 
\end{align}
\end{subequations}
{and we have set Boltzmann's constant $k_B = 1$ in the definition of the entropy.} 
The minimization of $\Omega$ with respect to $\hat D$ leads to the grand-canonical ensemble
\begin{align}
\hat D_{\rm gc} &= \frac{1}{Z_{\rm gc}}e^{-\beta \hat{H} + \alpha_p \hat{N}_p + \alpha_n \hat{N}_n} \, ,
\label{eq:density_op}
\end{align}
where $\beta = 1/T$, $\alpha_q = \beta \mu_q$, and $Z_{\rm gc}$ is the grand-canonical
partition function
\begin{align}
Z_{\rm gc} = \Tr  e^{-\beta \hat{H} + \alpha_p \hat{N}_p + \alpha_n \hat{N}_n }  \;.
\label{eq:partition}
\end{align}
The presence of $Z_{\rm gc}$  in Eq.~\eqref{eq:density_op} ensures
the normalization of the density operator, i.e., $\Tr \hat D_{\rm gc}=1$. 

The finite-temperature HF, HF+BCS and HFB approximations are obtained from the 
general variational principle for $\Omega$ using different forms for the density
operator $\hat{D}$. The chemical potentials are determined to reproduce 
 the given particle number $N_q$ for each species (protons and neutrons) 
on average, i.e., $\Tr (\hat D_{\rm gc} \hat N_q) = N_q$. 

\subsection{Hamiltonian and model space}

The CI shell model basis we employ is the many-particle space spanned by a set 
of spherical single-particle states characterized by their good quantum 
numbers {$n,l,j,m$, where $n$ is the principal radial quantum number, $l$ is the
orbital angular momentum,  and $j$ is the total angular momentum with 
z-projection $m$.} { While this set of orbitals usually originates from a 
physical one-body potential (e.g., harmonic oscillator or Woods-Saxon central 
potential plus spin-orbit interaction), the relevant quantities that describe 
the CI shell model are the single-particle energies, the two-body interaction 
matrix elements between pairs of the single-particle orbitals, and the matrix 
elements of relevant one-body operators such as the quadrupole operator. }

{In traditional shell-model applications, the chosen orbitals span the valence 
space outside of an inert core and only the valence nucleons are considered. 
For ``no-core" shell model calculations, there is no 
core, and all nucleons are included in the calculation.
Although the examples we provide here include an inert core, the code can in principle 
also be used for no-core calculations. We have, however, not 
included explicit strategies to improve the scaling of storage and CPU 
requirements with the size of the model space, and calculations might become 
impractical in very large no-core shell model spaces.  }

We use the proton-neutron formalism, in which protons and neutrons need not 
occupy the same valence spaces. In the following, we assume the single-particle
space to include $M_p$ proton orbitals and  $M_n$ neutron orbitals. 

The CI shell model Hamiltonian consists of a one-body part $\hat{H}^{(1)} $ and 
a two-body interaction. In general
\begin{align}
\hat{H} &=  \hat{H}^{(1)} + \hat{H}^{(2)}_{pp} + \hat{H}^{(2)}_{nn} + \hat{H}^{(2)}_{pn}  \, ,
\label{eq:hamil}
\end{align}
where

\begin{subequations}
\begin{align}
\hat{H}^{(1)} &=
  \sum_{i}^{M_p} \epsilon_{p,i}^{\rm SM} \hat{a}_{p,i}^{\dagger} \hat{a}_{p,i} 
+ \sum_{i}^{M_n} \epsilon_{n,i}^{\rm SM} \hat{a}_{n,i}^{\dagger} \hat{a}_{n,i} \, , \\
\hat{H}^{(2)}_{qq}&=
      \frac{1}{4} \sum_{ijkl}^{M_q} \bar{v}^{qq}_{ijkl} \, 
       \hat{a}_{q,i}^{\dagger} \hat{a}_{q,j}^{\dagger}
       \hat{a}_{q,l} \hat{a}_{q,k} \;\;\;   (q=p,n) \,, \\
\hat{H}^{(2)}_{pn}&=
       \sum_{ik}^{M_p} \sum_{jl}^{M_n} v^{pn}_{ijkl} \, 
       \hat{a}_{p,i}^{\dagger} \hat{a}_{n,j}^{\dagger}
       \hat{a}_{n,l} \hat{a}_{p,k} \, .
\end{align}
\end{subequations}
The operators $\hat{a}^{\dagger}, \hat{a}$ are the fermion creation 
and annihilation operators corresponding to the shell-model single-particle 
basis. The one-body Hamiltonian in Eq.~\eqref{eq:hamil} is characterized by  
single-particle energies $\epsilon^{\rm SM}_p, \epsilon^{\rm SM}_{n}$ and a
set of two-body matrix elements 
(TBMEs) $\bar{v}^{pp}, \bar{v}^{nn}$  and $v^{pn}$. The $pp$ and $nn$ matrix elements  
$\bar{v}^{pp}$ and $\bar{v}^{nn}$ are antisymmetrized, but the $pn$
matrix elements $v^{pn}$ are not. 

HF-SHELL accepts input in terms of the angular momentum coupled TBMEs. 
We express the uncoupled TBMEs in term of the 
coupled  matrix elements  with good angular momentum $JM$

\begin{subequations}
\begin{align}
\bar{v}^{qq}_{ijkl} &= \sqrt{1 + \delta_{ij}} \sqrt{1 + \delta_{kl}} \nonumber \\
 & \sum_{JM} 
          \left( j_i m_i j_j m_j | J M \right)
          \left( j_k m_k j_l  m_l | J M \right) \bar{v}^{J,qq}_{ijkl}\, ,
\label{eq:interaction_coupling} \\
v^{pn}_{ijkl} &=
   \sum_{JM} 
          \left( j_i m_i j_j  m_j | J M \right)
          \left( j_k m_k j_l  m_l | J M \right) v^{J,pn}_{ijkl} \, , 
\end{align}
\end{subequations}
where the $\left( j_i m_i  j_j m_j | J M \right)$ are 
Clebsch-Gordan coefficients. While the uncoupled matrix elements depend 
explicitly on the magnetic quantum numbers of the single-particle states, the 
coupled matrix elements are independent of them and of $M$. 
The good angular momentum TBMEs $ \bar{v}^{J,qq}$ and $v^{J,pn}$ 
are part of the input to the code. 
The formatting of the input regarding the
model space and the interaction are discussed in Sec.~\ref{sec:modelinput}.

While essentially all textbooks present the mean-field formulation for just one 
nucleon species, we will explicitly use both species in the  
 formalism such that the text follows more closely the
implementation in the code.

\subsection{The HF approximation}

The HF approximation is obtained using the variational principle for $\Omega$ 
when we assume a density matrix of the form
\begin{align}
\hat{D}_{\rm HF} = \frac{1}{Z_{\rm HF}^{0}} 
                \prod_{q=p,n} e^{-\beta \hat{K}_q + \alpha_q \hat{N}_q}  \, ,
\label{eq:HFansatz}
\end{align}
where $\hat{K}_q = \sum_{ij=1}^{M_q} K_{q,ij} \hat{a}^{\dagger}_{q,i} \hat{a}_{q,j}$
are one-body operators, and 
$Z^{\rm HF}_{0} = \prod_{q=p,n} \Tr  e^{-\beta \hat{K}_q + \alpha_q \hat{N}_q }$ 
is the corresponding partition function with $\alpha_q=\beta \mu_q$.  The one-body density 
matrix $\rho_q$ is defined 
by $\rho_{q,ij} = \Tr ( \hat{D} \hat{a}^{\dagger}_{q,j} \hat{a}_{q,i})$. 
Using Wick's theorem for statistical mixtures~\cite{Gaudin1960}, we find
\begin{align}
\rho_q &  = \frac{1}{1 + e^{\beta K_q - \alpha_q}}\;.
\label{eq:one-body} 
\end{align}
We can express the energy, entropy and particle numbers in Eqs.~(\ref{eq:E-S-N})
 in terms of the one-body density matrices $\rho_q$ 
\begin{subequations}
\begin{align}
E_{\rm HF} = & \sum_{q=p,n} \left[ \tr \left( \epsilon^{\rm SM}_q \rho_q \right)
            + \frac{1}{2} \tr \left( \Gamma_{q} \rho_{q} \right)\right]  \;, 
\label{eq:HF-E} \\
  S_{\rm HF} =  &  
       - \sum_{q=p,n} \left\{ \tr (\rho_q \ln \rho_q) + \tr[(1-\rho_q)\ln(1-\rho_q)]\right\}  \;, \\
  N_q = &  \tr \rho_q  \,,
\end{align}\label{eq:HF-E-S-N}
\end{subequations}
where $\tr$ indicates a trace over the single-particle space
of the relevant nucleon species, e.g., 
\begin{align}
\tr\left( \Gamma_{p} \rho_{p} \right) &=
                                \sum_{ij=1}^{M_p} \Gamma_{p,ij} \rho_{p,ji} \, .
\end{align}
The matrices $\Gamma_{p}$ and $\Gamma_{n}$ in Eq.~\eqref{eq:HF-E} are given 
by 
\begin{subequations}
\begin{align}
\Gamma_{p,ij} &= \sum_{kl=1}^{M_p}   \bar{v}^{pp}_{ikjl} \rho_{p,lk}   
                 + \sum_{kl=1}^{M_n}     v^{pn}_{ikjl}   \rho_{n,lk}
\, , \\ 
\Gamma_{n,ij} &= \sum_{kl=1}^{M_n} \bar{v}^{nn}_{ikjl} \rho_{n,lk} 
                 + \sum_{kl=1}^{M_p}  v^{pn}_{kilj} \rho_{p,lk}
\, .
\end{align}
\label{eq:Gamma}
\end{subequations}
Using the variational principle for $\Omega$ with respect to the 
density operator $\hat{D}_{\rm HF}$, or equivalently with respect to the 
matrix elements of $\hat{K}_q$, we find
\begin{align}
K_q &=  h_q \equiv \epsilon^{\rm SM}_q + \Gamma_q  \; , 
\label{eq:HF}
\end{align}
where $h_q$ is the single-particle HF Hamiltonian for $q=p,n$.
Combining Eqs.~\eqref{eq:one-body}  and \eqref{eq:HF}, we find the self-consistent finite-temperature HF equation 
\begin{align}
\rho_q &  = \frac{1}{1 + e^{\beta (\epsilon^{\rm SM}_q + \Gamma_q )   - \alpha_q}}\;.
\label{eq:FTHF}
\end{align} 

The formulation of the self-consistent problem can be simplified by 
transforming to the HF single-particle basis in which both $h_p$ and 
$h_n$ are diagonal with the HF single-particle energies $\epsilon_q$ along the diagonal. We call this basis the HF basis and denote 
the fermion creation and annihilation operators in this basis 
by $\hat{c}_q^{\dagger}, \hat{c}_q$. 
In the HF basis, the one-body density matrices 
are diagonal with occupation numbers $f_{q,i}$
\begin{align}
\rho_{q,ij} &=  \delta_{ij}f_{q,i} = 
            \delta_{ij} \frac{1}{1 + e^{\beta \epsilon_{q,i} -  \alpha_q }} \, ,
\label{eq:particleocc}
\end{align}
where $\epsilon_{q,i}$ are the HF single-particle energies. 
Eq.~\eqref{eq:particleocc} implies that $\rho_q^2 \not = \rho_q$, in contrast to the 
zero-temperature HF density matrix. This is since the nuclear configuration at 
finite  temperature is no longer a single Slater determinant but a statistical 
mixture $D_{\rm HF}$ of such determinants. 

The entropy can be easily calculated in the HF basis
\begin{align}
S_{\rm HF} &= - \sum_{q= p,n}
       \sum_{i=1}^{M_q} \left[ f_{q,i} \ln f_{q,i} + (1-f_{q,i})\ln (1-f_{q,i})\right] \, .
\end{align}
%
\subsection{The HFB approximation}
The HFB density matrix assumes a more general form than 
Eq.~\eqref{eq:HFansatz}
\begin{align}
\hat{D}_{\rm HFB} = \frac{1}{Z^0_{\rm HFB}} \prod_{q = p, n} e^{-\beta \hat{K}_q + \alpha_q \hat{N}_q } \, ,
\label{eq:HFBansatz}
\end{align} 
where $\hat{K}_q$ is given by the bilinear form

\begin{align}
\hat{K}_q 
&=
\frac{1}{2} 
\begin{pmatrix}
\hat{a}_q \\
\hat{a}_q^{\dagger}
\end{pmatrix}^{\dagger}
\begin{pmatrix}
K^{11}_q &  K^{20}_q \\
{-K^{20,*}_q} & -K^{11,{*}}_q
\end{pmatrix}
\begin{pmatrix}
\hat{a}_q \\
\hat{a}_q^{\dagger}
\end{pmatrix}
\nonumber \, ,\\
&=
\frac{1}{2} 
\begin{pmatrix}
\hat{a}_q \\
\hat{a}_q^{\dagger}
\end{pmatrix}^{\dagger}
\mathcal{K}_q
\begin{pmatrix}
\hat{a}_q \\
\hat{a}_q^{\dagger}
\end{pmatrix} \, .
\label{eq:K-operator} 
\end{align}
We define both the one-body density matrix $\rho_q$ and the anomalous density 
matrix $\kappa_q$
\begin{alignat}{2}
\rho_{q,ij}   &= \Tr \left( \hat{D}_{\rm HFB} \hat{a}^{\dagger}_j \hat{a}_i \right)  \,, & \;\;
\kappa_{q,ij} &= \Tr \left( \hat{D}_{\rm HFB} \hat{a}_j \hat{a}_i \right)  \, .
\end{alignat}
Together they form the generalized density matrices $\mathcal{R}_q$, each 
of dimension {$2M_q \times 2 M_q$}
\begin{align}
\mathcal{R}_q &= 
\begin{pmatrix}
\rho_q       & \kappa_q \\
-\kappa_{q}^* & 1 - \rho_{q}^* 
\end{pmatrix} \, .
\label{eq:R} 
\end{align}
Using Wick's theorem, ${\cal R}_q$  can be expressed in terms of the matrix 
representation ${\cal K}_q$ of the operator $\hat{K}_q$ 
(see Eq.~(\ref{eq:K-operator}))
\begin{align}
\mathcal{R}_{q}= \frac{1}{1 + e^{\beta \mathcal{K}_q - \alpha_q \mathcal{N}_q}} \, ,
\label{eq:R-K}
\end{align}
where
\begin{align}
\mathcal{N}_q &= 
\begin{pmatrix}
1 & 0 \\
0 & -1
\end{pmatrix} \, ,
\end{align}
is the matrix representation of the number operator $\hat N_q$.
We can express the total energy as a function of $\rho_q$ 
and $\kappa_q$ 
\begin{align}
E_{\rm HFB} &= \sum_{q=p,n}\left[ \tr \left( \epsilon^{\rm SM}_q \rho_q \right)
               +\frac{1}{2} \tr \left( \Gamma_q \rho_q \right)   +\frac{1}{2} \tr \left( \Delta_q \kappa^\dagger_q \right)  \right]
\, .
\end{align}
The HF potentials $\Gamma_{p}, \Gamma_{n}$ are given by Eqs.~\eqref{eq:Gamma} 
as in the HF case,  while the pairing potentials $\Delta_q$ are defined by
\begin{align}
\Delta_{q,ij}  &= \frac{1}{2} \sum_{kl=1}^{M_q}\bar{v}^{qq}_{ijkl} \kappa_{q,kl}  \, . 
\label{eq:gapdef}
\end{align}
The variation of $\Omega$ with respect to the matrix elements of $\mathcal{K}_p$
and $\mathcal{K}_n$ leads to the condition that
\begin{align}
\mathcal{K}_q &=
\begin{pmatrix}
h_q       &  \Delta_q \\
-\Delta^{*}_q & - h^*_q
\end{pmatrix} \, .
\label{eq:FTHFB}
\end{align}
Eq.~(\ref{eq:FTHFB}) together with Eqs.~(\ref{eq:R}) and (\ref{eq:R-K}) form the
HFB self-consistent equations for $\rho_q$ and $\kappa_q$. 

 We define the HFB 
Hamiltonian $\mathcal{H}_q$ to include the contribution of the chemical 
potential, i.e., $ \mathcal{H}_q = \mathcal{K}_q - \mu_q \mathcal{N}_q$~\cite{RingSchuck}. 
It is convenient to work in the quasi-particle basis that diagonalizes 
$\mathcal{H}_q$. For $k=1, \ldots, 2 M_q$, we have
\begin{align}
\mathcal{H}_q
\begin{pmatrix}
U_{q,k} \\
V_{q,k} 
\end{pmatrix}
&=
E^{\rm qp}_{q,k}
\begin{pmatrix}
U_{q,k} \\
V_{q,k} 
\end{pmatrix}
\label{eq:Hhfbdiag}\, ,
\end{align}
where $E^{\rm qp}_{q,k}$ are the quasi-particle energies and $U_{q,k}$, 
$V_{q,k}$ are column vectors of dimension $M_q$. The corresponding matrices 
$U_q$ and $V_q$ define a unitary Bogoliubov transformation $\mathcal{W}_q$ of 
dimension {$2M_q \times 2M_q$},  which transforms the
set of single-particle operators $\hat{a}_q^{\dagger}, \hat{a}_q$ 
into the set of quasi-particle operators $\hat{\beta}^{\dagger}_q, \hat{\beta}_q$
\begin{equation}
\begin{pmatrix}
\hat{\beta}_q         \\
\hat{\beta}_q^{\dagger} \\
\end{pmatrix}
=
\mathcal{W}_q
\begin{pmatrix}
\hat{a}_q           \\
\hat{a}_q^{\dagger} \\
\end{pmatrix}
=
\begin{pmatrix}
U_q^{\dagger} & V_q^{\dagger} \\
V^{T}_q & U^T_q \\
\end{pmatrix}
\begin{pmatrix}
\hat{a}_q           \\
\hat{a}_q^{\dagger} \\
\end{pmatrix}
\label{eq:BogoTransform} \, .
\end{equation}
This transformation brings the generalized density matrix in Eq.~\eqref{eq:R}
into a diagonal form
\begin{align}
\mathcal{R}_q &= 
\mathcal{W}_q
\begin{pmatrix}
f_q  & 0      \\
0  & 1 - f_q
\end{pmatrix}
\mathcal{W}_q^{\dagger} \, ,
\label{eq:diaggeneralized}
\end{align}
where $f_q$ is a diagonal matrix with the quasi-particle occupations along its 
diagonal
\begin{align}
f_{q,k} &=  \frac{1}{1 + e^{\beta E^{\rm qp}_{q,k}}} \, .
\label{eq:quasiparticleocc}
\end{align}
As in the HF case, non-vanishing quasi-particle occupations imply that the 
generalized density matrix is no longer a projector, i.e. 
$\mathcal{R}^2_q \not = \mathcal{R}_q$.  In contrast to the  zero-temperature HFB approximation,
the corresponding finite-temperature configuration is no 
longer described by the quasi-particle vacuum.

In the quasi-particle basis, the HFB entropy is given by an expression similar 
to the HF entropy 
\begin{align}
S_{\rm HFB} = - \sum_{q=p,n}\sum_{i=1}^{M_q} \left[ f_{q,i} \ln f_{q,i} + (1 - f_{q,i}) \ln (1-f_{q,i}) \right] \, ,
\end{align}
but the $f_{q,i}$ are now the quasi-particle occupations of 
Eq.~\eqref{eq:quasiparticleocc}.

The formalism simplifies when the HFB Hamiltonian is invariant under time-reversal symmetry. 
In that case, the quasi-particle states come in doubly 
degenerate pairs ($k,\bar{k}$) that are related through time-reversal, and 
the corresponding Bogoliubov matrix ${\cal W}_{q,T}$ can be expressed as a matrix of 
dimension {$M_q \times M_q$}~\cite{Fanto2017,RingSchuck}:
\begin{align}
\begin{pmatrix}
\hat{\beta}_{q,k} \\
\hat{\beta}^{\dagger}_{q,\bar{k}}
\end{pmatrix}
&=
\mathcal{W}^{\dagger}_{q,T}
\begin{pmatrix}
\hat{a}_{q,k} \\
\hat{a}^{\dagger}_{q,\bar{k}}
\end{pmatrix} \, ,  &&
\mathcal{W}_{q,T} &= 
\begin{pmatrix}
U_q & -V^*_q \\
V_q & U^*_q
\end{pmatrix} \, ,
\label{TR-symmetry}
\end{align}
where $k$ runs over only half the number of single-particle states 
($k=1,\ldots,M_q/2$). 

Finally, we remark on the concept of the canonical basis.  As is the case at 
zero-temperature, one can decompose the Bogoliubov transformation in 
Eq.~\eqref{eq:BogoTransform} following the Bloch-Messiah-Zumino 
theorem~\cite{Bloch1962,Zumino1962,RingSchuck}.  However, at finite temperature 
the matrices $\rho$ and $\kappa$ cannot be brought simultaneously into a 
diagonal form for $\rho$ and a canonical form for $\kappa$ 
by a single unitary transformation of the single-particle basis.
Thus, the notion of canonical basis does not generalize to the finite-temperature HFB
equations.

\subsection{The HF+BCS approximation}

The HF+BCS approximation consists of a further simplification of the HFB 
approximation for time-reversal invariant systems. In addition to time-reversal
symmetry, we assume that the interaction in the pairing 
channel acts mainly between time-reversal partners in the HF basis. 
In that basis we assume
\begin{align}
\Delta_{q,ij} &\approx \Delta_{q,i} \delta_{i \bar{j}} \, .
\end{align}
The Bogoliubov transformation can then be decomposed into a set of $M_q/2$ uncoupled 
transformations
\begin{equation}
\begin{pmatrix}
\hat{\beta}_{q,k}           \\
\hat{\beta}_{q,\bar{k}}^{\dagger} \\
\end{pmatrix}
=
\begin{pmatrix}
 u_{q,k}^* &           -v_{q,k}^* \\
 v_{q,k}   & \phantom{-}u_{q,k}  \\
\end{pmatrix}
\begin{pmatrix}
\hat{c}_{q,k}           \\
\hat{c}_{q,\bar{k}}^{\dagger} \\
\end{pmatrix}
\label{eq:BCSTransform} \, , 
\end{equation}
where the $u_k,v_k$ are now numbers. The diagonalization of the 
HFB Hamiltonian leads to
\begin{subequations}
\begin{align}
|u_{q,k}|^2 &= \frac{1}{2} \left( 1 + \frac{\epsilon_{q,k} - \mu_q}{E^{\rm qp}_{q,k}}\right) \, , 
\label{eq:u} \\
|v_{q,k}|^2 &= \frac{1}{2} \left( 1 - \frac{\epsilon_{q,k} - \mu_q}{E^{\rm qp}_{q,k}}\right) \, ,
\label{eq:v} 
\end{align}
\end{subequations}
with quasi-particle energies given by
\begin{align}
E^{\rm qp}_{q,k} = \sqrt{\left(\epsilon_{q,k} - \mu_q\right)^2  + \Delta_{q,k}^2} \, .
\label{eq:eqp}
\end{align}
The quasi-particle occupations retain the form of Eq.~\eqref{eq:quasiparticleocc}.
The matrices $\rho_q$ and $\kappa_q$ take a simple form in each of the 
two-dimensional subspaces composed of the time-reversed partners

\begin{alignat}{2}
\rho_{q} &= 
\begin{pmatrix}
\rho_{q,kk} & 0 \\
0                 & \rho_{q,\bar{k}\bar{k}}
\end{pmatrix} \, , \qquad
\kappa_q &=
\begin{pmatrix}
0                        &  \kappa_{q,k\bar{k}} \\
 \kappa_{q,\bar{k}k}  & 0 
\end{pmatrix} \, , 
\end{alignat}
with 
\begin{subequations}
\begin{align}
\rho_{q,kk}         &= \phantom{+}\rho_{q, \bar{k}\bar{k}} = \,  f_{q,k} + |v_{q,k}|^2 (1 - 2 f_{q,k}) \, , \\
\kappa_{q,k\bar{k}} &= -\kappa_{q,\bar{k}k}   = \, u_{q,k} v^*_{q,k}  (1 - 2 f_{q,k}) .
\end{align}
\end{subequations}
Together with the definition of the pairing gaps in Eq.~\eqref{eq:gapdef}, these
equations lead to finite-temperature gap equations
for each nucleon species~\cite{Goodman1981}
\begin{align}
\Delta_{q,i} &= -\frac{1}{2} \sum_{j=1}^{M_q/2} \bar{v}^{qq}_{i\bar{i}j\bar{j}} ( 1 - 2 f_{q,j}) \frac{\Delta_{q,j}}{E^{\rm qp}_{q,j}}  \,.
\label{eq:gapBCS}
\end{align}
The sum in Eq.~(\ref{eq:gapBCS}) ranges over half of the single-particle states
(such that the other half are their time-reversed partners). At zero 
temperature, the quasi-particle occupations vanish and Eq.~\eqref{eq:gapBCS}
reduces to the well known gap equations.

\subsection{Self-consistent symmetries}

If conserved symmetries are present, the number of variational parameters of the 
mean-field equations (the matrix elements of the HF or Bogoliubov 
transformation) can be significantly reduced, resulting in a lesser numerical effort.
The code assumes good parity and a conserved
z-signature $\hat{R}_z = e^{i \pi \hat{J}_z}$, as well as the reality of the
matrix elements of the HF and HFB transformations. We also assume time-reversal
symmetry, which implies that the mean-field description is suitable for even-even nuclei only. 
Finally, the proton-neutron formalism as presented here  
implies there is no mixing of proton and neutron single-particle orbitals.

\subsection{Iterative solutions of the mean-field equations}

The mean-field equations are nonlinear. In the HF approximation, the one-body density matrix 
$\rho$ can be obtained from the diagonalization of the single-particle 
Hamiltonian, which in turn depends on $\rho$. In the HFB 
approximation, the generalized density matrix $\mathcal{R}$ is the result of the
 diagonalization of $\mathcal{H}$, which in turn depends on $\mathcal{R}$ 
through the mean fields $\Gamma$ and the pairing gaps $\Delta$. 

For this reason, the self-consistent mean-field equations are solved by 
iterations. In HFB, starting from an initial guess for $\rho$ and $\kappa$,
 we construct the fields $\Gamma, \Delta$ and diagonalize the 
resulting Hamiltonian $\mathcal{H}$ to obtain new matrices
$\rho$ and $\kappa$. Several methods have 
been proposed in the literature. The most straightforward method of iterating 
and re-diagonalizing $\mathcal{H}$ which is constructed from
the improved $\rho$ and $\kappa$ often leads to convergence issues, as small 
changes in the matrix can give rise to large differences in the nuclear 
configuration.

Gradient methods~\cite{Davies1980,Egido1995,Robledo2011} 
evolve the configuration in a more gradual way, and are 
less susceptible to convergence issues. Such methods are particularly 
suitable for calculations in large model spaces since the iterations can be 
implemented without storing the full matrix $\mathcal{H}$. 
This advantage is not significant for relatively small model spaces such as those used in 
shell-model calculations, but the robustness of the self-consistency cycle
remains an attractive feature. Here we employ the heavy-ball optimization method 
of Ref.~\cite{Ryssens2019}, a simple extension of the original method of 
Ref.~\cite{Davies1980}, which we describe in Sec.~\ref{sec:numerical}.

\subsection{Mean-field partition functions}

It is not immediately apparent that the finite-temperature mean-field 
approximations satisfy the usual thermodynamical relations, which we refer to as
thermodynamic consistency. In this section we show that the finite-temperature
mean-field approximations are thermodynamically consistent. 

The starting point is the grand-canonical partition function, 
which can be obtained from the relation between the partition function and the 
grand-canonical potential,  $\ln Z = -\beta \Omega$. In the HF approximation, 
we find
\begin{align}
Z^{\rm HF}_{\rm gc} 
 &= e^{-\beta U^0_{\rm HF}} \prod_{q=p,n}
\prod_{i=1}^{M_q} \left[ 1 + e^{-\beta \epsilon_{q,i} + \alpha_q}\right] \, ,  \label{eq:Zhf} 
\end{align}
with 
\begin{align}
U^0_{\rm HF} =&  - \frac{1}{2} \sum_{q=p,n}\tr  \left( \Gamma_q \rho_q\right) \, .
\end{align}
Similarly, in the HFB approximation, we have
\begin{align}
Z^{\rm HFB}_{\rm gc} 
            &=   e^{-\beta U_{\rm HFB}^0} 
      \prod_{q=p,n} \prod_{i=1}^{M_q} \left( 1 + e^{-\beta E^{\rm qp}_{q,i}} \right) \, , 
\label{eq:Zgc_hfb}
\end{align}
with
\begin{align}
U_{\rm HFB}^0  =& - \frac{1}{2} \sum_{q=p,n} \tr \large( \Gamma_q \rho_q   \large)
                  - \frac{1}{2} \sum_{q=p,n} \tr \large( \Delta_q \kappa^{\dagger}_q \large) 
\, . 
\end{align}
The grand-canonical partition function in the HF+BCS approximation is 
formally identical to Eq.~\eqref{eq:Zgc_hfb}.

Thermodynamic consistency requires that the following relations hold 
\begin{align}
E^{\rm HF}    &= - \left. \frac{\partial \ln Z^{\rm HF}_{\rm gc}}{\partial \beta}\right|_{\alpha_p, \alpha_n} \,  ,
\label{eq:consistency1} \\
N_q^{\rm HF}  &= \phantom{-} \left. \frac{\partial \ln Z^{\rm HF}_{\rm gc}}{\partial \alpha_q}\right|_{\beta, \alpha_{q' \not= q}} \, ,
\label{eq:consistency2}
\end{align}
and similarly for the HFB and HF+BCS approximations. The validity of 
Eqs.~\eqref{eq:consistency1} and \eqref{eq:consistency2} is a nontrivial 
observation, since the single-particle Hamiltonian $\hat{h}$ depends on the 
temperature and chemical potential, as do the pairing gaps $\Delta$ in the HFB 
or HF+BCS approximations.

Following Appendix B of Ref.~\cite{Alhassid2016}, we prove
Eq.~\eqref{eq:consistency2} for protons in the HF approximation. 
The validity of Eq.~\eqref{eq:consistency1}, as well as the analogous equations
for the HFB and HF+BCS approximations can be proven along similar lines.
We rewrite the derivative in Eq.~\eqref{eq:consistency2} in the form
\begin{align}
\left. \frac{\partial \ln Z^{\rm HF}_{\rm gc}}{\partial \alpha_p}\right|_{\beta, \alpha_{n}}
&=  \left. - \frac{ \partial (\beta \Omega_{\rm HF})}{\partial \alpha_p}\right|_{\beta, \alpha_{n}} \nonumber \\
&=   N_p  - 
        \left.  \frac{ \partial (\beta \Omega_{\rm HF})}{\partial \hat{D}_{\rm HF}}\right|_{\beta, \alpha_{n}, \alpha_p}
        \left. \frac{ \partial \hat{D}_{\rm HF}}{\partial \alpha_p}\right|_{\beta, \alpha_{n}} \, .
        \label{eq:derivative}
\end{align}
The second term on the r.h.s.~of Eq.~\eqref{eq:derivative} vanishes since 
$\hat{D}^{\rm HF}$  is an extremum of the grand-canonical
potential $\Omega$. 

We emphasize the importance of the temperature-dependent constants $U^0$ in the 
mean-field approximations to the grand-canonical partition function, which implies
$Z_{\rm gc} \not= Z^0$ (where $Z^0$ is the partition function of the non-interacting particles). This is often overlooked in introductory texts, with 
the exception of Ref.~\cite{BlaizotRipka}. Without the inclusion $U^0$, 
the mean-field approximations would not satisfy Eqs.~\eqref{eq:consistency1}, 
\eqref{eq:consistency2} and their HFB and HF+BCS analogues.
\subsection{The zero-temperature limit}

We first discuss the zero-temperature limit in the presence of pairing 
correlations, i.e., we assume that a finite pairing gap exists for 
low temperatures in the HFB or HF+BCS approximations.  The quasi-particle energies are
all strictly positive and all quasi-particle occupations of Eq.~\eqref{eq:quasiparticleocc} vanish
as $T \rightarrow 0$. The generalized density matrix then assumes 
 its zero-temperature form
\begin{align}
\lim_{T \rightarrow 0} \mathcal{R}_q =
\begin{pmatrix}
0 & 0 \\
0 & 1
\end{pmatrix} \, .
\end{align}
In this limit, $\mathcal{R}_q$ is idempotent, and we recover the 
interpretation of the self-consistent mean-field solution as a pure state -- 
the vacuum of a set of quasi-particle operators determined by the Bogoliubov 
transformation.

The $T\rightarrow 0$ limit is less straightforward when pairing correlations
collapse at low temperatures or in the HF approximation; 
see Ref.~\cite{Duguet2020a} and Appendix B of Ref.~\cite{Levit1984}. When the highest HF single-particle 
level is degenerate and partially occupied (open shell), the $T\to 0$ limit is a
 statistical mixture, in which all the degenerate orbitals of the highest HF 
single-particle level have the same occupation probability. 
Thus, we do not recover the single Slater determinant description of the $T=0$ HF approximation.
This zero-temperature limit is known as the uniform filling approximation. When 
the highest HF single-particle level is fully occupied (closed shell), the 
statistical mixture reduces to a projector and the usual $T=0$ HF description
 is obtained.

\section{Canonical partition functions and calculation of state densities}
\label{sec:levelden}
The finite-temperature mean-field approximations discussed above
are all formulated in the grand-canonical ensemble. For a finite nucleus 
however, both the proton and neutron numbers are fixed, and it is necessary 
to reduce the formulation to the canonical ensemble. One could incorporate 
particle-number projection operators $\hat{P}_n$ and $\hat{P}_p$
into the ansatz for the density operator $\hat{D}$ and apply the variational principle for the canonical free energy.
This procedure is known as variation after projection (VAP)~\cite{Esebbag1993}.
The method is however impractical at finite temperature because the entropy term
depends on the logarithm of a particle-projected density operator.  This 
approach has been implemented only in relatively small model spaces
where direct diagonalization is feasible, see, e.g., Ref.~\cite{Gambacurta2012}.

In contrast,  projection after variation, in which particle-number 
projection operators are applied to the self-consistent grand-canonical 
mean-field solution is more tractable. Here we summarize the 
ensemble-reduction method of Ref.\cite{Fanto2017}, which starts from the 
particle-number projected partition function. We only include the key formulas 
that are implemented in the code, and refer the interested reader to the 
original reference for more details. We assume time-reversal symmetry; for a 
generalization to the case where time-reversal symmetry may be broken see 
Ref.~\cite{Fanto2017a}.

In the HF approximation, the number-projected canonical partition function 
$Z^{\rm HF}_{\rm c} $ is given by
\begin{align}
Z^{\rm HF}_{\rm c} &= e^{-\beta U_{\rm HF}^0} 
\Tr \left( \prod_{q = p,n} \hat{P}_q e^{-\beta \hat{h}_q + \alpha_q \hat{N}_q} \right) \,.
\label{eq:startcanonical}
\end{align}
The particle-number projection can be carried out explicitly using a discrete 
Fourier transform, leading to
\begin{subequations}
\begin{align}
Z^{\rm HF}_{\rm c} &= e^{-\beta U_{\rm HF}^0} \prod_{q =p,n} 
C_q 
\left( 
 \sum_{\ell=1}^{M_q} e^{-i\phi_{q,\ell} N_q} \zeta^{\rm HF}_{q,\ell} 
\right) 
\, , 
\label{eq:ZcHF}\\
C_q &= \frac{e^{-\alpha_q N_q}}{M_q}\, , \\
\zeta_{q,\ell}^{\rm HF} &= 
\prod_{k=1}^{M_q} \Big( 1 + e^{-\beta \epsilon_{q,k} + \alpha_q + i \phi_{q,\ell}}\Big) \, ,
\end{align}
\end{subequations}
where the quadrature gauge angles are $\phi_{q,\ell} = \frac{2 \pi \ell}{M_q}$. 

In the HFB approximation, we assume time-reversal symmetry for which the 
Bogoliubov transformation has the form of Eq.~(\ref{TR-symmetry}).  
We also define the auxiliary matrices  $\mathcal{N}_q$ and $\mathcal{E}_q$
\begin{align}
\mathcal{N}_q = \begin{pmatrix}
1 & 0       \\
0          & -1
\end{pmatrix} \, ,
&&
\mathcal{E}_q = \begin{pmatrix}
E^{\rm qp}_q & 0       \\
0          & -E^{\rm qp}_q
\end{pmatrix} \, , 
\end{align}
where $E^{\rm qp}_q = E^{\rm qp}_1, \ldots E^{\rm qp}_{M_q/2}$ are the 
quasiparticle energies of every pair. Note that both these matrices are {of dimension 
 $M_q \times M_q$}. The canonical partition function is then given by
\begin{subequations}
\begin{align}
Z^{\rm HFB}_{\rm c}   
 &= e^{-\beta U_{\rm HFB}^0  } \prod_{q=p,n} C_q 
                       \left( \sum_{\ell=1}^{M_q} e^{-i \phi_{q,\ell} N_q} \zeta^{\rm HFB}_{q,\ell} \right) \, , 
\label{eq:ZcHFB}\\
C_q &=  e^{- \frac{\beta}{2}\tr \left( h_q - \mu_q \right)} \frac{e^{-\alpha_q N_q}}{M_q} \, , \\
\zeta^{\rm HFB}_{q,\ell} &= (-1)^{\ell} 
\text{det} \Big(1 + \mathcal{W}^{\dagger}_{T,q} e^{i \phi_{q,\ell} \mathcal{N}_q} \mathcal{W}_{T,q} e^{-\beta \mathcal{E}_q}\Big) 
\label{eq:problematicdeterminant} \, ,
\end{align}
\end{subequations}
where $\mathcal{W}_{T,q}$ is the reduced Bogoliubov transformation matrix
of Eq.~\eqref{TR-symmetry}.

In the HF+BCS approximation, this expression can be rewritten using the 
$u_q, v_q$ parameters of the BCS transformation in the HF basis
\begin{subequations}

\begin{align}
Z^{\rm BCS}_{\rm c} &   
=e^{-\beta U_{\rm BCS}^0} \prod_{q=p,n}
C_q 
\left(
\sum_{\ell=1}^{M_q} e^{-i \phi_{q,\ell} N_q} \zeta^{\rm BCS}_{q,\ell} 
\right)\, , 
\label{eq:ZcBCS}
\\
C_q &
=  e^{- \frac{\beta}{2}\tr \left( h_q - \mu_q \right)}\frac{e^{-\alpha_q N_q}}{M_q}  
\, ,\\
\zeta^{\rm BCS}_{q,\ell} &= (-1)^{\ell} 
\prod_{k=1}^{M_q/2} \Big[
  e^{+\beta E^{\rm qp}_{q,k}}   \big(|u_{q,k}|^2 + e^{2 i \phi_{q,\ell}} |v_{q,k}|^2\big) \nonumber \\
&+ e^{- \beta E^{\rm qp}_{q,k}}  \big(|v_{q,k}|^2 + e^{2 i\phi_{q,\ell}}|u_{q,k}|^2\big)  + 2 e^{ i \phi_{q,\ell}}  \Big] \, ,
\label{eq:zetaBCS}
\end{align}
\end{subequations}
where now the product in Eq.~\eqref{eq:zetaBCS} ranges over 
only half the single-particle states. 

Individual terms in Eqs.~\eqref{eq:ZcHF}, \eqref{eq:ZcHFB} and \eqref{eq:ZcBCS} 
can be very small or very large at low temperatures. For this reason, the 
code calculates the logarithm of the relevant canonical partition 
function. In particular, we use the  
QDR-decomposition technique discussed in Appendix B of Ref.~\cite{Fanto2017} to
calculate the determinant in Eq.~\eqref{eq:problematicdeterminant} 
in a numerically stable way.

The state density is related to the canonical partition function $Z_c$ through
an inverse Laplace transform, which we evaluate using the saddle-point 
approximation\footnote{{Also known as the stationary phase approximation.}}~\cite{Mathews1970,Chaichian2018}
\begin{align}
\rho(E, N_p, N_n) &= \frac{1}{2 \pi i} \int_{-i \infty}^{+i \infty} d \beta' e^{\beta' E} Z_{\rm c} \nonumber \\
&\approx \left( 2 \pi \left| \frac{\partial E}{\partial \beta}\right| \right)^{-1/2} e^{\rm S_{\rm c}(\beta)} \, .
\label{eq:statedensity}
\end{align}
In Eq.~\eqref{eq:statedensity}, the saddle-point condition determines the 
inverse temperature $\beta$ as a function of the canonical energy $E_c$ as 
\begin{align}
E = -\frac{\partial \ln Z_{\rm c}}{\partial \beta} \equiv E_{\rm c}(\beta).
\label{eq:dZdB}
\end{align}
The canonical entropy $S_{\rm c}$ can then be obtained from the canonical partition function and
energy
\begin{align}
S_{\rm c} = \ln Z_{\rm c} + \beta E_{\rm c} \, .
\label{eq:dEdB}
\end{align}
We emphasize that Eq.~\eqref{eq:statedensity} is an expression for the 
nuclear \emph{state} density, which counts all many-body states, including 
the $(2 J + 1)$-fold degeneracy of a level with spin $J$. In contrast, 
the  nuclear \emph{level} density counts each degenerate level with spin $J$ 
only once.  In the spin-cutoff 
model~\cite{Ericson1960},  these densities are related by
\begin{align}
\rho_{\rm level}(E) &= \frac{1}{\sqrt{2 \pi \sigma^2}} \rho_{\rm state}(E) \, , 
\label{eq:spincut}
\end{align}
where $\sigma$ is the spin-cutoff parameter.

The most efficient way to calculate the state density in 
practice is to (i) perform mean-field calculations at many different 
inverse temperatures $\beta$ and  (ii) use a finite-difference formula 
to evaluate the canonical  energy and entropy. The code performs the relevant Fourier sum, either 
Eq.~\eqref{eq:ZcHFB}, Eq.~\eqref{eq:ZcHF} or Eq.~\eqref{eq:ZcBCS}, after the 
iterative process is completed.
Calculations for several temperatures can be carried out  in a single execution of 
 the code, and the output is given in tabulated form.
To obtain the state or level
density using Eqs.~\eqref{eq:dEdB}, \eqref{eq:dZdB} and \eqref{eq:statedensity} 
we include an auxiliary script \texttt{level\_densities.py} which processes 
this tabulated output as discussed in 
Sec.~\ref{sec:auxiliaries}.

\section{Observables}
\label{sec:observables}

Other than the total energy, entropy and particle numbers, the code outputs 
information on several other observables as discussed below. 
\paragraph{Decomposition of the energy}
We decompose the total energy into three parts
\begin{align}
E_{\rm tot} = E_{\rm sp} + E_{\rm 2b} + E_{\rm pair} \, , 
\end{align}
where we define the single-particle, two-body and pairing energies by
\begin{subequations}
\begin{align}
E_{\rm sp}   &= \tr \Big( \epsilon^{\rm SM}_p \rho_p \Big) + \tr \Big( \epsilon^{\rm SM}_n \rho_n \Big) \, ,   \\
E_{\rm 2b}   &=   \frac{1}{2} \tr \Big( \Gamma_p \rho_p \Big) + \frac{1}{2} \tr \Big( \Gamma_n \rho_n \Big)  \, , \\
E_{\rm pair} &=  \frac{1}{2} \tr \Big( \Delta_p \kappa^{\dagger}_p \Big)
                 +\frac{1}{2} \tr \Big( \Delta_n \kappa^{\dagger}_n \Big) \, .
\label{eq:}
\end{align}
\end{subequations}

\paragraph{Quadrupole moments}
The code utilizes several different parameterizations of quadrupole deformation
to characterize the nuclear shape following Ref.~\cite{Ryssens2015a}.
The intrinsic quadrupole deformation operators 
$\hat{Q}_{2m}$ ($m=-2,\ldots,2$) are defined as
\begin{align}
\hat{Q}_{2m}  &=  \sqrt{\frac{16\pi}{5}} \sum_i r_i^2 Y_{2m}(\theta_i, \phi_i) \, ,
\end{align}
where the sum is over both protons and neutrons and 
$\hat{Y}_{2m}(\theta_i, \phi_i)$ is a spherical harmonic. 
Because of the self-consistent 
symmetries assumed in the code, only $\langle \hat{Q}_{20} \rangle_T$ and  
$\Re \langle \hat{Q}_{22} \rangle_T = \Re \langle \hat{Q}_{2-2} \rangle_T $ 
do not vanish. In order to calculate the 
single-particle matrix elements of these operators, the code requires an input  
 file which contains the (reduced)
matrix elements of $r^2$ for the model space, see Sec.~\ref{sec:input}. 
If such file is not provided, the code will calculate the matrix elements for 
harmonic oscillator basis functions; see, e.g., Ch.~6 in Ref.~\cite{SuhonenBook}. 

 For triaxial shapes, it is perhaps more intuitive to use the 
$q,\gamma$ variables defined as
\begin{subequations}
\begin{align}
q
& =  \sqrt{\langle \hat{Q}_{20}\rangle_T^2 + 2 \langle \hat{Q}_{22} \rangle_T^2} \, , \\
\gamma &= \text{atan2} \left( \sqrt{2} \, \langle \hat{Q}_{22} \rangle_T , \langle \hat{Q}_{20} \rangle_T \right) \, .
\end{align}
\end{subequations}
Another set of variables, $q_1$ and $q_2$, is useful to parametrize in an 
unambiguous way individual sextants of the full $(q, \gamma)$ plane. They are 
given by 
\begin{subequations}
\begin{align}
q_{1} & =  q\cos\gamma - \frac{1}{\sqrt{3}}q\sin\gamma \, , \\
q_{2} & =  \frac{2}{\sqrt{3}}q\sin\gamma \, .
 \label{q2}
\end{align}
\end{subequations}
The code also outputs the variance of the total quadrupole moment
\begin{align}
\text{var} (Q^2) = 
\sum_{m=-2}^{2} \langle \hat{Q}^*_{2 m} \hat{Q}_{2 m}\rangle_T
            -   \langle \hat{Q}^*_{2 m}\rangle_T \langle \hat{Q}_{2 m}\rangle_T\, .
\end{align}
\paragraph{Average pairing gaps}

Unlike the case of a simple seniority interaction, there is no unique definition of a
pairing gap for a general two-body interaction. The pairing gap matrices 
$\Delta^q$ (which simplify in the BCS approximation) are orbital 
dependent. In order to extract meaningful information about the strength of 
pairing correlations, we define average pairing gaps.  Assuming all 
matrices to be real and time-reversal to be conserved, we define
\begin{subequations}
\begin{align}
\langle v^2 \Delta \rangle_q &=  \frac{\sum_{ij=1}^{M_q} \rho_{q,ij} \Delta_{q,j\bar{i}}}{\tr \rho_q}     \, ,  \label{eq:v2delta_HFB}\\ 
\langle u v \Delta \rangle_q &=  \frac{\sum_{ij=1}^{M_q} \kappa_{q,ij} \Delta_{q,ji}}{ \sum_{ij=1}^{M_q}  |\kappa_{q,ij}|} \, .  \label{eq:uvdelta_HFB} 
\end{align}
\end{subequations}
In the BCS case, these expressions reduce to the definitions of Ref.~\cite{Bender2000}
\footnote{At $T=0$, we can use these definitions also in the canonical HFB basis. However, for $T\not= 0$, a canonical basis generally does not exist.}
\begin{subequations}
\begin{align}
\langle v^2 \Delta \rangle_q &=  \frac{\sum_{k=1}^{M_q} v^2_k \Delta_{k}}{ \sum_{k=1}^{M_q}  v^2_k}     \, ,  \label{eq:v2delta_BCS}\\ 
\langle u v \Delta \rangle_q &=  \frac{\sum_{k=1}^{M_q} u_k v_k \Delta_{k}}{ \sum_{k=1}^{M_q}  u_k v_k} \, .  \label{eq:uvdelta_BCS} 
\end{align}
\end{subequations}
Eqs.~\eqref{eq:v2delta_HFB} and \eqref{eq:v2delta_BCS} are not equivalent
to Eqs.~\eqref{eq:uvdelta_HFB} and \eqref{eq:uvdelta_BCS}.
In the shell-model single-particles space we use they 
typically differ by 5\% or less.
\paragraph{Belyaev moment of inertia }

The code calculates and outputs the Belyaev moments of inertia
$\mathcal{I}_{\mu \mu}$ around the three main axes $\mu = x,y,z$~\cite{Belyaev1961} .
We denote by $J_{\mu, q, kl}$, the matrix elements of $\hat{J}_{\mu}$
for nucleon species $q$ in the HF basis, and assume time-reversal
symmetry.
In the HF approximation, the moments of inertia are given by
\footnote{When pairing correlations vanish, this
formula reduces to the Inglis moment of inertia~\cite{Inglis1954}.}
\begin{align}
\mathcal{I}^{\rm HF}_{q,\mu \mu} & = \sum_{k,l=1}^{M_q/2} 
\left( 2 |J_{\mu, q, kl}|^2   
              + 2 |J_{\mu, q, \bar{k}l}|^2 \right)\frac{f_{q,l} - f_{q,k}}{\epsilon_{q,k} - \epsilon_{q,l}} \, .
\label{eq:IHF}
\end{align}
For terms in Eq.~\eqref{eq:IHF} involving degenerate single-particle states,
 we take the appropriate limit 
\begin{align}
\lim_{\epsilon_{q,k} \rightarrow \epsilon_{q,l}} 
\frac{f_{q,l} - f_{q,k}}{\epsilon_{q,k} - \epsilon_{q,l}}
&= \beta e^{\beta \epsilon_{q,k}} f_{q,k}^2 \, .
\label{eq:limit}
\end{align}

In the BCS approximation,  assuming $u_{q}, v_{q}$ to be real, we have~\cite{Alhassid2005}
\begin{subequations}
\begin{align}
\mathcal{I}^{\rm BCS}_{q,\mu \mu} & = \sum_{k,l=1}^{M_q/2} 
\left( 2 |J_{\mu, q, kl}|^2   
              + 2 |J_{\mu, q, \bar{k}l}|^2 \right) W^{\rm BCS}_{q,kl} \, ,
\label{eq:IBCS}
\\
W^{\rm BCS}_{q,kl} &=  (u_{q,k} u_{q,l} + v_{q,k} v_{q,l})^2 \, \frac{f_{q,l} - f_{q,k}}{E^{\rm qp}_{q,k} - E^{\rm qp}_{q,l}} \nonumber \\
                   &  +(u_k v_l - v_k u_l)^2 \,  \frac{1 - f_{q,k} - f_{q,l}}{E^{\rm qp}_{q,k} + E^{\rm qp}_{q,l}}\, . 
\label{eq:Wbcs}
\end{align}
\end{subequations}
In the HFB approximation, we have in the quasi-particle basis
\begin{align}
\mathcal{I}^{\rm HFB}_{q,\mu \mu} =  2 \sum_{k,l=1}^{M_q} \Bigg[ &
  \left( |J^{20}_{\mu,q, kl}|^2 + |J^{20}_{\mu, q, k\bar{l}}|^2 \right) \frac{1 - f_{q,k} - f_{q,l}}{E^{\rm qp}_{q,k} + E^{ \rm qp}_{q,l}} \nonumber \\
 +&  \left( |J^{11}_{\mu, q, kl}|^2 + |J^{11}_{\mu, q, k\bar{l}}|^2 \right) \frac{f_{q,l} - f_{q,k}}{E^{\rm qp}_{q,k} - E^{\rm qp}_{q,l}} \Bigg] \, , 
\end{align}
where the indices range over half of the quasi-particle space, and
the matrices $J^{20}_{\mu,q}$ and $J^{11}_{\mu,q}$ are given 
by~\cite{RingSchuck}
\begin{subequations}
\begin{align}
J^{20}_{\mu,q} &= U^{\dagger}_q J_{\mu,q} V^{*}_q - V^{\dagger}_q J^{T}_{\mu,q} U^{*}_q \, , \\
J^{11}_{\mu,q} &= U^{\dagger}_q J_{\mu,q} U_q  - V^{\dagger}_q J^{T}_{\mu,q} V_q \, . 
\end{align}
\end{subequations}
For degenerate quasi-particle energies in the BCS and 
HFB approximations, we can apply formulas analogous to Eq.~\eqref{eq:limit}.
\section{{Methods of solution}}
\label{sec:numerical}

In order to solve the self-consistent mean-field equations, HF-SHELL uses
methods similar to those discussed in Refs.~\cite{Ryssens2019} and 
\cite{Ryssens2015a}. We will summarize these methods and present the relevant 
equations. A schematic diagram of the code is shown in Fig.~\ref{fig:diag}.
\subsection{Heavy-ball algorithm}
The heavy-ball algorithm was recently  proposed~\cite{Ryssens2019} to 
iteratively diagonalize the single-particle Hamiltonian $\hat{h}$. 
We assume a set of orthonormal single-particle wavefunctions 
$|\psi^{(i)}_{q,k} \rangle$ ($q=p,n$ and $k=1, \ldots, M_q$), given at iteration $i$. 
For the next iteration $i+1$, we define 
\begin{equation}
| \phi^{(i+1)}_{q,k} \rangle = | \psi^{(i)}_{q,k} \rangle 
     - \alpha \left(  \hat{h}_q^{(i)}   -   \epsilon_{q,k}^{(i)}\right)| 
\psi^{(i)}_{q,k} \rangle  + \mu  |\delta \psi^{(i)}_{q,k} \rangle \label{eq:update} 
 \end{equation}
where $\hat h^{(i)}_q$ is the single-particle Hamiltonian for species $q$
calculated from the one-body matrix density matrix at iteration $i$, and 
the quantities $\epsilon^{(i)}_{q,k}$ and $|\delta \psi^{(i)}_{q,k} \rangle$
 are given by
\begin{subequations}
\begin{align}
\epsilon_{q,k}^{(i)} & = \langle \psi^{(i)}_{q,k} | \hat{h}^{(i)}_q |\psi^{(i)}_{q,k} \rangle \, ,\\
|\delta \psi^{(i)}_{q,k} \rangle & =  | \psi^{(i)}_{q,k} \rangle  - | \psi^{(i-1)}_{q,k} \rangle \, .
\end{align}
\end{subequations}
The parameters $\alpha$ and $\mu$ \footnote{These purely numerical parameters 
are not to be confused with the chemical potential $\mu_q$ and $\alpha_q = \beta \mu_q$
that were introduced in Section~\ref{sec:mean-field}.
} 
in Eq.~(\ref{eq:update}) are called the 
stepsize and momentum, respectively. When $\mu = 0$, this algorithm reduces to 
the gradient descent method\footnote{Also known as the imaginary time-step method.}
 of Ref.~\cite{Davies1980}. For $\mu \not = 0$, the algorithm can achieve 
convergence faster than the gradient descent method, at the moderate cost of 
storing the $|\delta \psi{^{(i)}}_{q,k}\rangle$ and virtually no extra CPU 
cost. Finally,  the wavefunctions
 $|\phi_k^{(i+1)} \rangle$ are not orthonormal, and we employ a Gram-Schmidt 
procedure to construct an orthonormal basis  $|\psi^{(i+1)}_{q,k} \rangle$, 
from which we can start a new iteration.

The convergence of the algorithm is sensitive to the values of $\alpha$
and $\mu$~\cite{Ryssens2019}.  Their values can be input by the user, but we 
recommend not doing so in most cases. In the absence of such input, the code uses the 
following values:
\begin{subequations}
\begin{align}
\alpha &=  0.9  \frac{4}{\left(\sqrt{\delta\epsilon} + \sqrt{0.1}\right)^2} \, , &
\, \\
\mu    &= \left( \frac{\sqrt{\delta\epsilon/0.1} - 1}{\sqrt{\delta\epsilon/0.1} + 1}\right)^2  ,
\label{eq:evol}
\end{align}
\end{subequations}
where $\delta \epsilon$ (measured in MeV) is the difference between the largest 
and smallest {estimates $\epsilon^{(i)}_{q,k}$} of 
eigenvalues of the current single-particle Hamiltonian. The 
speed-up of the heavy-ball algorithm with respect to ordinary gradient descent 
is modest.  Speed-ups are often smaller than a factor of two, but can be  
up to a factor of four for heavier nuclei such as the lanthanides discussed below 
in Sec.~\ref{sec:examples}.  The reason why this performance gain is not a large as those 
reported in Ref.~\cite{Ryssens2019} is the difference in the span of the
single-particle spectra. Shell model Hamiltonians do not typically
have single-particle spectra for which $\delta \epsilon$ is larger than a 
few MeV.

In special circumstances Eqs.~\eqref{eq:evol} might not work well. 
One such case occurs when calculating quadrupole constrained surfaces
for light nuclei at low temperatures in the absence of pairing correlations 
(or in the HF approximation). The Lagrange parameters for the 
quadrupole constraints can then take very large values and the single-particle 
spectrum is effectively dominated by them.  We advise the user in such cases to
 input by hand values for $\alpha$ and $\mu$.

\subsection{Two-basis method for the HFB problem}

In the HF approximation, the heavy-ball algorithm is the main 
component of the self-consistency cycle. For the HF+BCS approximation, this 
algorithm needs to be supplemented by the calculation of the $u_q,v_q$ pairs
in Eqs.~\eqref{eq:u} and \eqref{eq:v}, but this does not add significant 
complexity. For the HFB approximation, however, it is not immediately obvious 
how to employ the heavy-ball algorithm to diagonalize the HFB Hamiltonian. 

Dedicated gradient-based algorithms for the HFB approximation 
exist~\cite{Egido1995,Robledo2011}, but we employ a simpler
approach: the two-basis method. This approach consists of rewriting 
Eq.~\eqref{eq:Hhfbdiag} as
\begin{align}
\tilde{\mathcal{H}}_q 
\begin{pmatrix}
\tilde{U}_{q,k} \\
\tilde{V}_{q,k} \\
\end{pmatrix}
&=
\begin{pmatrix}
\epsilon_{q} - \mu_q & \tilde{\Delta}_{q} \\
- \tilde{\Delta}^{*}_{q}  & - \epsilon_{q} + \mu_q \\
\end{pmatrix}
\begin{pmatrix}
\tilde{U}_{q,k} \\
\tilde{V}_{q,k} \\
\end{pmatrix} \, ,
\label{eq:H_HFBalt}
\end{align}
where the tildes indicate matrices calculated in the
HF basis, in which the matrices $\epsilon_{q}-\mu_q$ are diagonal. {The  
vectors $(\tilde{U}_{q,k}, \tilde{V}_{q,k})^T$ are then obtained at every 
iteration by diagonalizing the HFB Hamiltonian $\tilde{\mathcal{H}_q}$.}

The two-basis method was originally proposed in Ref.~\cite{Gall1994} to simplify 
HFB calculations in large model spaces. In HF-SHELL, we explicitly store the
entire HF basis, rendering the method exact.
Our  reason to adopt this method is to make the treatment of the 
HFB approximation as similar as possible to the simpler HF and HF+BCS 
approximations, thus making the code more transparent.

\subsection{Quadrupole constraints}
\label{sec:quadconstraints}
It is often of interest to investigate the properties of a self-consistent 
mean-field solution as a function of various collective variables such as the 
nuclear quadrupole moment. HF-SHELL allows the user to constrain this quantity 
in the optimization process. We use the method of Lagrange multipliers, in which
 the grand potential potential $\Omega$ is replaced by
\begin{align}
\Omega \rightarrow  \Omega - \lambda_{20} \langle \hat{Q}_{20} \rangle_T
                           - \lambda_{22} \langle \hat{Q}_{22} \rangle_T \, .
\label{eq:constrained}
\end{align}
Here $\lambda_{20}$ and $\lambda_{22}$ are Lagrange multipliers conjugate to  the
quadrupole moments $\hat{Q}_{20}$ and $\hat{Q}_{22}$. This effectively
introduces an external potential field in the single-particle Hamiltonian
\begin{align}
\hat{h}_q \rightarrow \hat{h}_q - \lambda_{20} \hat{Q}_{20,q} - \lambda_{22} \hat{Q}_{22,q} \, ,
\end{align}
but otherwise does not affect the equations in 
Sec.~\ref{sec:mean-field}. 

The code allows for two different types of quadrupole constraints. The first 
is  the linear constraint type, in which the Lagrange multipliers are kept fixed
throughout the iterations to predefined values. It is not very practical for
generating free energy surfaces since a regular discretization of the 
Lagrange parameters will not result in a regular discretization of the 
quadrupole deformations of the mean-field solutions~\cite{Staszczak2010}. 
The second method is the Augmented Lagrangian approach, which allows the user 
to specify the desired quadrupole deformations $Q_{20}^{\rm target}$ and 
$Q_{22}^{\rm target}$. The method is based on an iterative method that adjust the
Lagrange parameters. At iteration $i+1$, we obtain new Lagrange parameters 
based on the current quadrupole deformation~\cite{Staszczak2010}
\begin{align}
\lambda_{2 m}^{(i+1)} &= \lambda_{2 m}^{(i)}  
      + 2 C_{m} \left( \langle \hat{Q}_{2 m} \rangle_T^{(i)} - Q_{2 m}^{\rm target} \right) \, ,
\label{eq:ALMupdate}
\end{align}
where $C_{m}$ are parameters, and we initialize 
$\lambda_{2 m}^{(0)} = 0$. If the parameters $C_{m}$ are left unspecified by the user
(which we recommend), HF-SHELL uses reasonable estimates that work in 
essentially every situation we have encountered. In addition to the update of 
Eq.~\eqref{eq:ALMupdate}, the code employs a corrective gradient scheme to 
force the heavy-ball update to favor the constraints. The 
details of this algorithm will be presented elsewhere~\cite{RyssensUnpub}.

A similar method could be easily implemented to constrain the 
expectation value of any one-body operator, provided it 
respects the self-consistent symmetries assumed by the code. 

\subsection{Convergence}

For convergence, we require the following two conditions 
to hold when proceeding from iteration $i-1$ to $i$
\begin{subequations}
\begin{align}
\left| \frac{ E_{\rm tot}^{(i)} - E_{\rm tot}^{(i-1)}}{E_{\rm tot}^{(i)}} \right| 
\leq \text{e\_prec} \, , \\
\sum_{m = 0,2}\left| \langle \hat Q_{2m} \rangle^{(i)}_T - \langle \hat Q_{2m} \rangle_T^{(i-1)} \right| \leq \text{q\_prec} \, , 
\end{align}
\end{subequations}
where e\_prec and q\_prec are input parameters that determine the precision of 
the solution. 

\subsection{Density mixing}
In cases where convergence of the self-consistency cycle is problematic, the code
offers the possibility to slow down the update of the density and anomalous
density matrix by using a simple linear mixing procedure. Instead of using directly
the calculated $\rho_q$ and $\kappa_q$ at any given iteration, the code constructs
a mixed density and anomalous density at iteration $i$ 
\begin{subequations}
\begin{align}
\rho^{(i)}_q   = \alpha_{\rm mix} \rho^{(i)}_q[\psi]  + (1- \alpha_{\rm mix})\rho^{(i-1)}_q  
\label{eq:mix:rho}
\, , \\
\kappa^{(i)}_q = \alpha_{\rm mix} \kappa^{(i)}_q[\psi] + (1- \alpha_{\rm mix})\kappa^{(i-1)}_q \, , 
\label{eq:mix:kappa}
\end{align}
\end{subequations}
where $\alpha^{\rm mix}$ is a numerical parameter between 0 and 1. Here we denote
by $\rho^{(i)}[\psi], \kappa^{(i)}[\psi]$ the density and anomalous density
constructed directly from the single-particle wavefunctions and the Bogoliubov or
HF transformation currently in storage. Due to the limited size of both the
model space and the heavy-ball steps, we found 
that density mixing is needed only in exceptional cases.

\section{Examples}
\label{sec:examples}
\begin{figure}
\includegraphics[width=.53\textwidth]{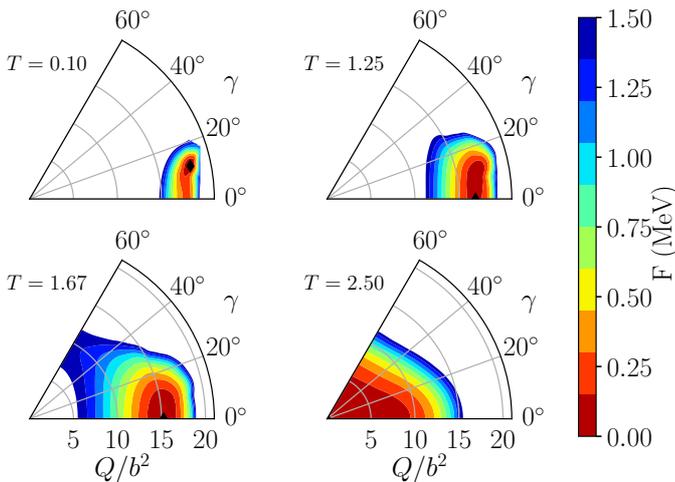}
\caption{Free energy surfaces of $^{24}$Mg with the USDB interaction as a function
         of $Q$ (in units of the square of the oscillator length $b$) and 
         $\gamma$ for four different values of the temperature (in units of 
         MeV).  The free energy is measured with respect to its global minimum 
         {(indicated by a
         black diamond)} at the given temperature.
          }
\label{fig:mg24_transition}
\end{figure}

\subsection{$^{24}$Mg in the $sd$ shell}

As a first example, we study a light nucleus using the well-known USDB effective
interaction~\cite{Brown2006} for $sd$-shell nuclei. As reported in 
Ref.~\cite{Gao2015}, almost all of the even-even $sd$-shell nuclei exhibit 
axially deformed mean-field minima. $^{24}$Mg is an exception, with a mean-field
deformation characterized by $\gamma = 12^{\circ}$. In 
Fig.~\ref{fig:mg24_transition} we show the free energy 
$F = E - TS$ as a function of the quadrupole deformation $Q$ 
(measured in units of $b^2$ with $b$ being the oscillator's radius)
and the angle $\gamma$ for four different values of the temperature. 
We observe two shape transitions: at low temperatures the minimum 
is triaxial with $\gamma = 12^{\circ}$, but around $T \sim 1.25$ MeV there is a shape
transition to an axial prolate configuration. At even higher temperatures, the 
prolate deformation gets washed out by thermal fluctuations and the free energy 
surface becomes flatter with a spherical mean-field minimum. We note that the
HF, HF+BCS, and HFB surfaces for this nucleus are all similar since 
pairing correlations are weak and vanish for almost all configurations at all 
temperatures.

\subsection{Lanthanide nuclei: $^{144}$Nd and $^{162}$Dy}

As an example of a larger model space, we study pairing and shape 
transitions in two lanthanide nuclei, $^{144}$Nd and $^{162}$Dy, which are to 
date among the heaviest studied with the shell model Monte Carlo (SMMC) 
method~\cite{Alhassid2008,Ozen2013}. Particle-number projected mean-field 
state densities in this mass region were benchmarked against SMMC level 
densities in Ref.~\cite{Alhassid2016}.

For these lanthanide nuclei, the single-particle model space consists of 40 
proton orbitals  ($0g_{7/2}$, $1d_{5/2}$, $1d_{3/2}$, $2s_{1/2}$, $0h_{11/2}$, 
and $1f_{7/2}$) and 66 neutron orbitals ($0h_{11/2}$, $0h_{9/2}$, $1f_{7/2}$, 
$1f_{5/2}$, $2p_{3/2}$, $2p_{1/2}$, $0i_{13/2}$, and $1g_{9/2}$). 
These orbitals form the valence space for protons and neutrons outside  
an inert $^{120}$Sn core.  
The single-particle energies and quadrupole matrix elements are
obtained from a Wood-Saxon potential with spin-orbit interaction using the 
parameters of Ref.~\cite{Bohr1998}. We use the effective interactions of 
Refs.~\cite{Alhassid2008,Ozen2013}.

\begin{figure}
\centering
\includegraphics[width=.45\textwidth]{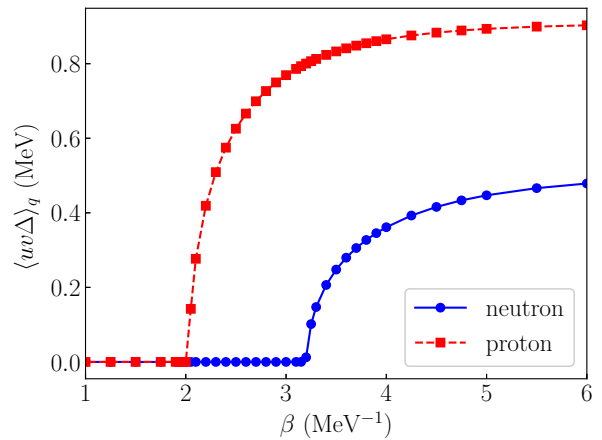}
\caption{Average pairing gap $\langle u v \Delta \rangle_q$ of $^{144}$Nd as a 
         function of the inverse temperature $\beta$ for protons (red squares) and neutrons (blue circles).}
\label{fig:Nd144}
\end{figure}

$^{144}$Nd is an example of a spherical nucleus with a strong pairing 
condensate. Fig.~\ref{fig:Nd144} shows the average 
pairing gap $\langle uv\Delta \rangle_q$ of Eq.~\eqref{eq:uvdelta_HFB} as a 
function of the inverse temperature $\beta$ for neutrons (blue circles) and 
protons (red squares). 
Pairing correlations vanish above a critical temperature of about $0.50$ MeV 
for protons and $0.31$ MeV for neutrons. 

$^{162}$Dy is a well-deformed nucleus with relatively weak pairing correlations.
In Table~\ref{tab:Dy162} we list {various physical quantities calculated 
at the zero-temperature mean-field minimum.} 
 While the difference in total binding energy between HF and HFB
solutions is less than $150$ keV and the difference in total quadrupole moment 
is very small, pairing correlations contribute in HFB more than $1.5$ MeV of 
binding energy (all due to the neutrons). Pairing correlations also have a 
significant effect on the Belyaev moment of inertia, whose HFB value is 
about 60\% of its HF value. 

\begin{table}
\centering
\begin{tabular}{l|S[table-format=6.3]S[table-format=6.3]}
$^{162}$Dy  & {HF} & {HFB} \\
\hline
\hline
$E_{\rm sp} $     & -407.598& -405.883  \\
$E_{\rm two-body}$&   35.817 & 35.534  \\
$E_{\rm pairing}$ &    0.000 & -1.560  \\
$E_{\rm tot}$     & -371.781 &-371.909    \\
$\langle Q_{20}\rangle$       &  653.508 & 652.353  \\
$\mathcal{I}_{xx}$&   85.798 & 54.799 \\
\hline
\end{tabular}
\caption{Comparison of zero-temperature quantities (energies, quadrupole moment $\langle Q_{20}\rangle$ and 
         Belyaev moment of inertia $\mathcal{I}_{xx}$ around an axis perpendicular to the symmetry axis) in $^{162}$Dy 
         for the HF and HFB solutions. All energies are in units of MeV, 
         $\mathcal{I}_{xx}$ is in units of $\hbar^{2}$MeV$^{-1}$ and the quadrupole 
         moment is in units of fm$^2$.}
\label{tab:Dy162}
\end{table}

\begin{figure}
\includegraphics[width=.45\textwidth]{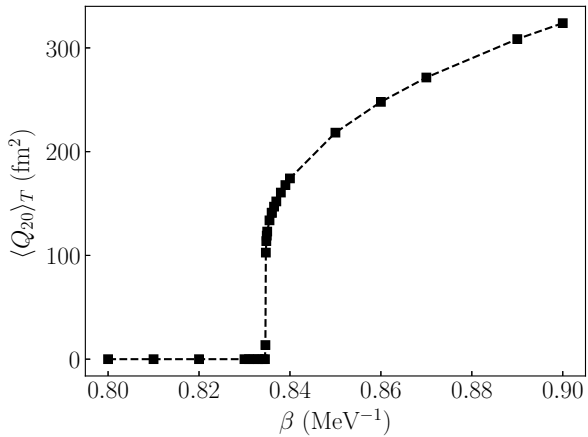}
\caption{The quadrupole moment $\langle \hat{Q}_{20}\rangle_T$  of $^{162}$Dy as
         a function of inverse temperature $\beta$ in the vicinity of the shape transition. }
\label{fig:Dy162}
\end{figure}

In Fig.~\ref{fig:Dy162}, we show the quadrupole moment $\langle Q_{20}\rangle_T$ 
of $^{162}$Dy as a function of the inverse temperature $\beta$, in the vicinity
of the shape transition temperature of $T = 1.2$ MeV. In this regime the HF, BCS
and HFB results are identical, as the temperatures shown are above the critical 
temperature of the pairing transition. In Fig. \ref{fig:Dy162}  we have used 
the same scales as in Fig.~1 of Ref.~\cite{Bertsch2016} to demonstrate the 
agreement between both codes.

\subsection{Nuclear state density of $^{162}$Dy}
\begin{figure}
\includegraphics[width=.45\textwidth]{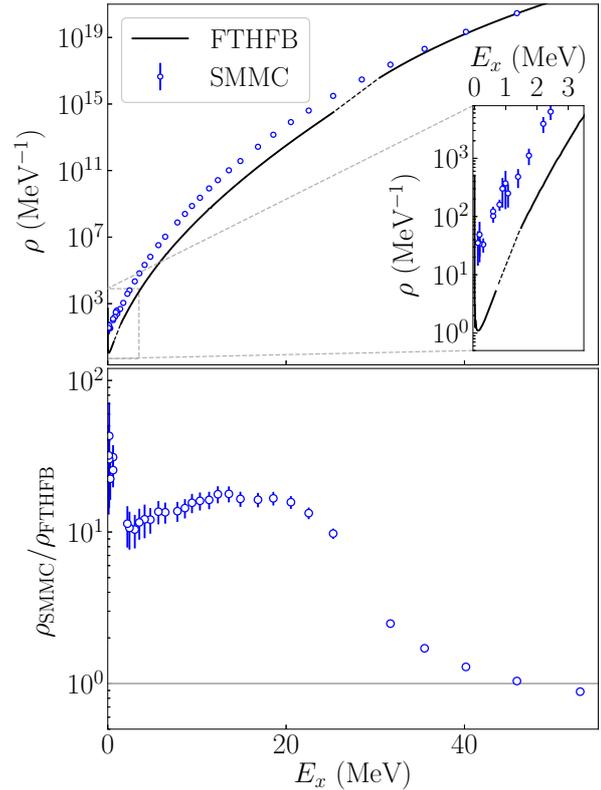}
\caption{ {Top panel: the} state density $\rho$ of $^{162}$Dy as a function of 
          excitation energy $E_x$, as calculated from HF-SHELL (solid black lines)
          and SMMC  (blue circles). 
          {Bottom panel: the ratio of the SMMC state density to the HFB state density.
          The light gray line describes a ratio of 1.
          Discontinuities associated with the neutron pairing phase 
          transition and the shape phase transition are visible in both panels,
          and are shown in the top panel by dashed lines.
           The SMMC results were taken from~\cite{Alhassid2016}.} }
\label{fig:rhoDy162}
\end{figure}

Here we demonstrate the mean-field calculation of state densities. {In the top
panel of Fig.~\ref{fig:rhoDy162} we compare the HFB state density of $^{162}$Dy 
obtained with HF-SHELL (black line) with the SMMC state density 
(blue circles)~\cite{Alhassid2016}.  In the bottom panel we show the ratio of 
the SMMC density to the HF-SHELL density.} Similar results for this nucleus were
 discussed in Refs.~\cite{Alhassid2016,Fanto2017}, in which the HF mean-field 
approximation was used. 

The {deficiencies} of the mean-field approach as discussed in these references are 
visible {in both panels of Fig.~\ref{fig:rhoDy162}:  (i) the mean-field 
results exhibit two discontinuities associated with the neutron pairing 
phase transition around an excitation energy of $0.7$ MeV and the shape phase 
transition around $25$ MeV (shown by dashed lines).  These transitions are 
absent in the exact results; (ii) the mean-field results underestimate 
significantly the SMMC state density below the pairing phase transition, which 
can be traced to  an unphysical negative entropy (not shown) and is an artifact
of performing projection after variation calculations~\cite{Fanto2017}; and 
(iii) the mean-field density underestimates the exact SMMC density by roughly a 
constant factor of 12 between $\sim 2$ MeV and the shape transition excitation 
energy. This can be explained by the effects of rotational collectivity: the 
exact state density includes all rotational bands, while the mean-field
results count only the deformed intrinsic band-heads~\cite{Alhassid2016}.  
The feature that the rotational enhancement is constant across a wide range of 
excitation energies is in line with more phenomenological 
models~\cite{Bjornholm73}, and was also observed in other rare-earth
nuclei~\cite{Ozen2013}. This rotational enhancement eventually disappears above 
the shape transition energy when the mean-field solutions conserve rotational 
symmetry.}

\section{Using the code }
\label{sec:usage}

\subsection{Compilation and execution}

{The code is written in Fortran 2003. Most of the code is backward compatible 
with Fortran 95 compilers, but we have employed procedure pointers in several 
places, which were introduced in the 2003 standard. To the best of our 
knowledge, all modern compilers implement this feature.}

We provide a Makefile with the source code. To compile the code 
execute the command 
\begin{verbatim}
make CXX=gfortran
\end{verbatim}
in the source code directory. The flag \textrm{CXX} 
determines the compiler used, and defaults to gfortran when left unspecified.
The code relies on the availability of a LAPACK distribution, which should be 
linked properly. 

\begin{figure}
\includegraphics[width=.45\textwidth]{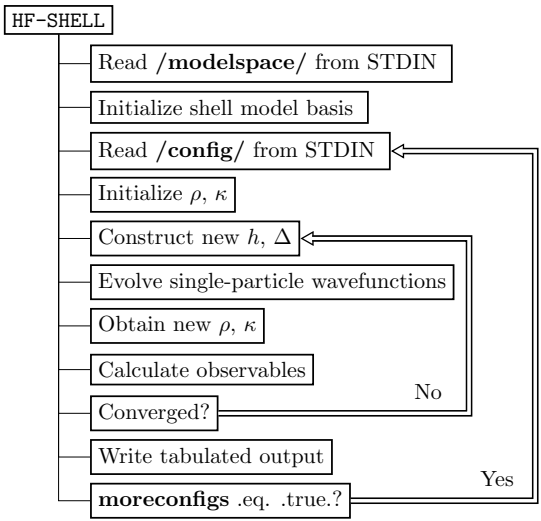}
\caption{A schematic diagram demonstrating the general logic and structure of 
         the code.}
\label{fig:diag}
\end{figure}

User input is read from STDIN. To run the code, execute 
\begin{verbatim}
./hf_shell.exe < input.data
\end{verbatim}
where \texttt{input.data} contains all run options for the code. It  
consists of two Fortran namelists, \texttt{/modelspace/} and \texttt{/config/}. 
We will describe the contents of these namelists in more detail in 
Sec.~\ref{sec:input}. The second of these namelists can be repeated by the 
user any number of times by setting the logical variable \textsf{moreconfigs=.true.}. 
A schematic diagram of the logic and structure of the code is shown in 
Fig.~\ref{fig:diag}.

The user should also provide a file describing the single-particle CI shell model 
space, as well as a file with the single-particle energies and the 
TBMEs of the effective interaction. Finally, the user has the option of 
providing a file that contains the reduced matrix elements of the quadrupole 
operator. 

\subsection{Input of CI shell-model space and Hamiltonian}
\label{sec:modelinput}

The code requires the user to provide two files, one that defines the 
single-particle orbitals and a second that provides the single-particle energies
and TBMEs. The format follows the conventions of the proton-neutron 
formalism of the SMMC~\cite{Alhassid2008} and 
BigStick~\cite{Johnson2013,Johnson2018} codes. This convention also matches that
of the mean-field code of Ref.~\cite{Bertsch2016}. For more information on the 
formatting of these files, see the examples which are included with the code 
distribution. 

{Many shell-model interactions in the literature employ a scaling with 
   mass number. If requested by the user, the code scales the TBMEs as follows}
{
\begin{align}
\bar{v}_{ijkl}^{qq} \rightarrow 
                  \left(\frac{A_{\rm ref}}{A} \right)^x \bar{v}_{ijkl}^{qq} \, ,
\label{eq:massscaling}
\end{align}}
{where $A$ is the mass number, $A_{\rm ref}$ is a reference mass and 
   $x$ is a real number. Both $A_{\rm ref}$ and $x$ are
  parameters of the interaction. A similar scaling is applied to 
  the $pn$ matrix elements.}

\subsection{Run options}
\label{sec:input}

A valid input file for the code consists of a namelist 
\texttt{/modelspace/} and one or more instances of the namelist 
\texttt{/config/}.  The first determines the single-particle model space and 
effective interaction, as well as some other general aspects of the calculation.
The \texttt{/config/} namelist, contains more specific options that the user 
might vary in a systematic fashion, such as the inverse temperature $\beta$ or 
the value of the quadrupole constraints. For this reason, this namelist can be 
repeated as often as needed so the code can explore varied parameters in a 
single run. 

\subsubsection{Options of the \texttt{/modelspace/} namelist}
\vspace{3 mm}

\begin{center}
\begin{tabular}{l l c}
\hline
 Parameter &  Type & Default value \\
\hline
\multicolumn{3}{c}{{Model space and interaction}}\\
\hline
    \texttt{spsfile}    & character &  `pn.sps'  \\
    \texttt{interfile}  & character &  `test.int'\\
    \texttt{qfile}      & character &  `r2.red'  \\
    \texttt{outfile}    & character &  `run.out' \\
    \texttt{TBME\_A}     & integer   &   -1 \\
    \texttt{TBME\_Aref} & integer   &   -1 \\
    \texttt{TBME\_x}    & real      &   -1.0 \\
\hline
\multicolumn{3}{c}{{Evolution and convergence}}\\
\hline
    \texttt{ptype}      & character & `HF' \\
    \texttt{maxiter}    & integer   &  200  \\
    \texttt{printiter}  & integer   &  10  \\
    \texttt{e\_prec}    & real      &  $10^{-9}$\\
    \texttt{q\_prec}    & real      &  $10^{-3}$\\
\hline
\end{tabular}
\end{center}

\textbf{Model space and interaction}

\begin{itemize}
\item \texttt{spsfile}: filename containing information on the
        single-particle model space. 
\item \texttt{interfile}: filename containing the single-particle energies
       and TBMEs of the CI shell-model Hamiltonian.
\item \texttt{qfile}: filename containing the reduced matrix elements of the 
       quadrupole operator. If blank, the code
       constructs matrix elements using harmonic oscillator wavefunctions, 
       see, e.g., Ch.~6 in Ref.~\cite{SuhonenBook}.
\item \texttt{outfile}: filename for the tabulated output.
\item \texttt{TBME\_A, TBME\_Aref, TBME\_x}: parameters of the mass scaling 
      of the TBMEs, Eq.~\eqref{eq:massscaling}. The scaling is disabled when 
      \texttt{TBME\_A} takes negative values.
\end{itemize}

\textbf{Evolution and convergence}
\begin{itemize}
\item \texttt{ptype}:  Type of mean-field approximation. The options are `HF', `BCS' or `HFB'.
\item \texttt{maxiter}: Maximal number of heavy-ball iterations.
\item \texttt{printiter}: Number of iterations between two consecutive 
                          full printouts.
\item \texttt{e\_prec}: Convergence parameter for the total energy.                 
\item \texttt{q\_prec}: Convergence parameter for the quadrupole moments
                        in units of fm$^2$.
\end{itemize}

\subsubsection{ Options of the \texttt{/config/} namelist}
\vspace{3 mm}

\begin{center}
\begin{tabular}{llc}
\hline
 Parameter & Type & Default value \\
\hline
\multicolumn{3}{c}{ General information}\\
\hline 
\texttt{  stepsize      }& real    &  0.0   \\
\texttt{  momentum      }& real    &  0.0   \\
\texttt{  inversetemp   }& real    & 32.0   \\
\texttt{  protons       }& integer &  6     \\
\texttt{  neutrons      }& integer &  4     \\
\texttt{  denmix        }& real    & 1.0    \\
\texttt{  moreconfigs   }& logical & .false.\\
\hline
\multicolumn{3}{c}{ Quadrupole constraints}\\
\hline
\texttt{  Q20target     }& real    & 0.0 \\
\texttt{  Q22target     }& real    & 0.0 \\
\texttt{  q1target      }& real    & 0.0\\
\texttt{  q2target      }& real    & 0.0\\
\texttt{  Q20c          }& real    & 0.0 \\
\texttt{  Q22c          }& real    & 0.0 \\
\texttt{  lambda20      }& real    & 0.0 \\
\texttt{  lambda22      }& real    & 0.0\\
\texttt{  constrainttype}& integer & 0  \\
\texttt{  constraintiter}& integer & 0 \\
\hline
\end{tabular}
\end{center}
\vspace{3 mm}
\textbf{General information}
\begin{itemize}
\item \texttt{stepsize, momentum}: parameters  $\alpha$ and $\mu$ of the heavy-ball update, 
                                   Eq.~\eqref{eq:update}. When not read from 
                                   input, the code uses the values of Eqs.~\eqref{eq:evol}.
                                  \item \texttt{inversetemp}: Inverse temperature $\beta$ in units of MeV$^{-1}$.
                           A negative value will result in a 
                            zero-temperature calculation. 
\item \texttt{protons/neutrons}: Number of protons/neutrons.
\item \texttt{denmix} : mixing parameter $\alpha_{\rm mix}$ for the density, 
                        see Eqs.~\eqref{eq:mix:rho} and Eqs.~\eqref{eq:mix:kappa}.
\item \texttt{more\_configs}: Logical variable indicating whether there will be more 
                             occurrences of the \textsf{/config/} namelist after
                             the current one.
\end{itemize}
\textbf{Quadrupole constraints}
\begin{itemize}
\item \texttt{Q20/22target}: target values for quadratic quadrupole constraints, in units of fm$^2$.
\item \texttt{q1/2target}: alternative way of specifying target values, using 
                           the $q_1, q_2$ convention, in units of fm$^2$.
\item \texttt{Q20/22c}: values of $C_{m}$ in Eq.~\eqref{eq:ALMupdate} for the quadrupole constraints in units of fm$^{-4}$ MeV.  When set to
                          zero, the code will use an estimate.
\item \texttt{lambda20/22}: Lagrange multipliers for linear \\ quadrupole  
                             constraints in units of MeV fm$^{-2}$.
\item \texttt{constrainttype}: Specifies the type of quadrupole constraints:
                               (0) no constraints,
                               (1) linear constraints,
                               (2) quadratic constraints.
\item \textsf{constraintiter}: If non-zero, the quadrupole constraints are only 
                                active for this number of iterations.
\end{itemize}

\subsection{Output}

HF-SHELL prints to STDOUT a large amount of information, 
such as the evolution of the iterative process as well as
 the complete HF basis and the properties of the 
quasi-particles in the HF+BCS or HFB approximations. 
The code also produces a  tabulated output file that contains the final values 
of various observables.
The filename of this output is determined by the input parameter \texttt{outfile}.
Every row of this file corresponds to one calculation, i.e.,  one 
occurrence of the namelist \texttt{/config/}. This file is particularly 
practical for further processing of a large number of sequential calculations, e.g.,  for constructing an energy surface or 
for calculating the state density. Table~\ref{tab:output} contains
the quantities that are tabulated in this file and their respective column numbers. 

\begin{table}
\begin{tabular}{lll}
\hline
Column  & Quantity  & Units\\
\hline
   1 & Run index\\
   2 & Proton number \\
   3 & Neutron number \\
   4 & Inverse temperature            & MeV$^{-1}$\\
   5 & Energy $E$                     & MeV\\
   6 & Free energy $F$                & MeV\\
   7 & Entropy $S$                    &        \\ 
   8 & $\langle \hat{Q}_{20} \rangle_T$ & fm$^2$ \\     
   9 & $\langle \hat{Q}_{22} \rangle_T$ & fm$^2$ \\    
  10 & Var(Q)                         & fm$^4$ \\    
  11 & $\ln Z_{\rm gc}$ & \\
  12 & $ \ln Z_{\rm c}$    & \\
  13-14-15 & $\langle \hat{J}^2_{x/y/z} \rangle_T$  & $\hbar^2$ \\
  16-17-18 & $\mathcal{I}_{xx/yy/zz}$ & $\hbar^2$ MeV$^{-1}$ \\
  19-20 & $\langle uv \Delta \rangle_{p/n}$ & MeV \\
  21-22 & $\langle v^2 \Delta \rangle_{p/n}$ & MeV \\
\hline
\end{tabular} 
\caption{Content of the tabulated output file.}
\label{tab:output}
\end{table}
\subsection{{Choice of initial configuration and sequential calculations}}  

{
When performing HF+BCS or HFB calculations, the code initializes at the first
iteration with an educated, non-zero guess for the pairing gaps.
Furthermore, the code uses the shell-model single-particle states as the  
initial Hartree-Fock basis so the initial mean-field configuration is spherically symmetric. For a 
Hartree-Fock calculation (or when the pairing gap vanishes), the code 
constructs the initial state by filling up the orbitals with the available number
of nucleons. Unless both nucleon numbers correspond to closed-shell 
configurations, the resulting confguration will not be spherically symmetric.}

{
In the majority of cases, the initial mean-field configuration is  
spherically symmetric. Without further input by the user, this symmetry 
will be conserved throughout the iterations.  For a nucleus with a deformed 
global minimum this spherical solution describes a saddle-point in the 
deformation surface. In order to break the spherical symmetry of the initial 
configuration, we recommend the use of the keyword \texttt{constraintiter}. By
setting some initial constraints for a small number of iterations, the code 
evolves towards a deformed solution at the start of the calculation. When these
iterations are completed, the constraint is turned off and the code is free to 
find the mean-field minimum, unconstrained by spherical symmetry. We recommend 
the user to set such initial constraints for both 
$\langle \hat{Q}_{20}\rangle_T$ and 
$\langle \hat{Q}_{22}\rangle_T$  (to also break axial symmetry).}

{
When several mean-field calculations are carried out sequentially 
(as signaled by occurrences of the namelist \texttt{/config/}), the code uses
the final mean-field configuration of the previous calculation as the starting 
point of the new calculation. This choice is particularly useful when 
calculating various observables as a function of temperature.}

\subsection{Auxiliary script: \texttt{level\_densities.py}}
\label{sec:auxiliaries}

For convenience, we also provide a Python script for the 
calculation of projected mean-field densities as described in 
Sec.~\ref{sec:levelden}. 
The python script \texttt{level\_densities.py}  assumes a working installation of 
Numpy~\cite{vanderWalt2011}. It includes a Python function
\begin{verbatim}
def generate_level_densities(fname, s2=[]):
\end{verbatim}
which employs a first-order finite-difference formula to evaluate the 
derivative in Eq.~\eqref{eq:dZdB}.
The first argument of the  function is the filename of an HF-SHELL tabulated 
output file produced by a run  
with different values of the inverse temperature. It returns a tuple of Numpy 
arrays, indexed by the inverse temperatures found in the HF-SHELL output

\begin{verbatim}
return (E_c, S_c, rho_state, rho_level)
\end{verbatim}
where \texttt{E\_c} is the canonical energy, \texttt{S\_c} is the canonical
entropy and \texttt{rho\_state} is the nuclear state density.
\texttt{rho\_level}  is the nuclear level density, obtained from the 
state density using Eq.~\eqref{eq:spincut} within the spin-cutoff model.
The second (optional) argument \texttt{s2} is $\sigma^2$, the square of the 
spin-cutoff parameter $\sigma$ used to obtain the level density. When this 
argument is not specified, the spin-cutoff parameter is taken from the 
mean-field calculation using~\cite{Alhassid2005}
\begin{align}
\sigma^2 &= \langle J_z^2 \rangle \, . 
\label{eq:sigma}
\end{align}
%

\section{Conclusion and outlook}\label{sec:conclusion}

{
We presented the code HF-SHELL, which can be used to explore
a variety of mean-field approximations for CI shell-model Hamiltonians, both at 
zero and finite temperature. It is particularly suitable for calculating
nuclear state densities in a mean-field approximation.} 

{The code requires little memory and its execution is typically very fast, making  
it a useful tool to supplement calculations that use other many-body methods. 
Possible future applications of this code  include (i) the benchmarking of more 
computationally intensive many-body methods; (ii)  the construction of effective 
CI shell-model interactions through fits of certain observables to experimental 
data, which are often very costly to undertake with more advanced many-body 
methods; and (iii)  the construction of reference states as a starting point 
for, e.g.,  the generator coordinate method, symmetry restoration methods, and 
Bogoliubov perturbation theory.
}

\begin{acknowledgements}
We thank P. Fanto for useful discussions and for comments on the manuscript. 
We also thank M. Bender for useful discussions, and in 
particular for pointing out the advantages of Brents algorithm for
root finding, and providing an example subroutine. We thank 
P. Stevenson for providing an example of a Fortran module for the calculation 
of Clebsch-Gordan coefficients.
This work was supported in part by the U.S. DOE grant No.~DE-SC0019521. 
\end{acknowledgements}

\bibliographystyle{spphys}       
\bibliography{hf_shell.bib}   

\end{document}